\documentclass[12pt]{article}

\usepackage{fullpage}
\usepackage{amssymb}
\usepackage{amsmath}
\usepackage{hyperref}
\usepackage{tensor}

\usepackage{caption}
\usepackage{subcaption}
\usepackage{tikz-cd}
\usepackage{tikz}
\usetikzlibrary{arrows}
\usepackage{tikz-3dplot}
\usepackage{pgfplots}
\usepgfplotslibrary{fillbetween}

\numberwithin{equation}{section}

\newcommand{\eps}{\epsilon}

\newcommand{\db}{\bar{\partial}}
\newcommand{\qb}{\bar{q}}

\newcommand{\G}{{\Gamma}}

\renewcommand{\k}{\kappa}
\renewcommand{\L}{{\mathcal L}}
\renewcommand{\O}{{\mathcal O}}
\newcommand{\N}{{\mathcal N}}

\newcommand{\w}{{\omega}}

\newcommand{\halfs}{\tfrac{1}{2}}
\newcommand{\C}{\mathbb{C}}
\renewcommand{\H}{\mathbb{H}}
\newcommand{\R}{\mathbb{R}}

\newcommand{\CP}{\mathbb{C}P}

\newcommand{\Z}{\mathbb{Z}}
\renewcommand{\S}{\mathcal{S}}
\newcommand{\U}{{\mathcal{U}}}
\DeclareMathOperator{\Tr}{Tr}
\DeclareMathOperator{\ind}{ind}
\let\Im\relax
\DeclareMathOperator{\Im}{Im}
\let\Re\relax
\DeclareMathOperator{\Re}{Re}
\DeclareMathOperator{\sgn}{sgn}
\newcommand{\bra}{\langle}
\newcommand{\ket}{\rangle}
\newcommand{\nn}{\nonumber}
\newcommand{\iso}{\cong}
\newcommand{\qvec}{{\underline{q}}}

\newcommand{\vq}{{\vec{q}}}
\newcommand{\vz}{{\vec{z}}}
\newcommand{\vw}{{\vec{w}}}
\newcommand{\zb}{{\bar{z}}}
\newcommand{\wb}{{\bar{w}}}

\begin{document}

\thispagestyle{empty}

\smallskip
\begin{center} \LARGE

{\bf 7D supersymmetric Yang-Mills
\\ on a 3-Sasakian manifold
} \\[12mm] \normalsize
{\bf Andreas Roc\'en} \\[8mm]
{\small \it Department of Mathematics, Uppsala University,\\
Box 480, SE-75106 Uppsala, Sweden.} \\[4mm]
\href{mailto:andreas.rocen@math.uu.se}{andreas.rocen@math.uu.se}
\medskip

\end{center}

\vspace{7mm}

\begin{abstract}
\noindent
In this paper we study 7D maximally supersymmetric Yang-Mills on a specific 3-Sasakian manifold that is the total space of an $SO(3)$-bundle over $\mathbb{C}P^2$. The novelty of this example is that the manifold is not a \emph{toric} Sasaki-Einstein manifold.
The hyperk\"ahler cone of this manifold is a Swann bundle with hypertoric symmetry and this allows us to calculate the perturbative part of the partition function of the theory. The result is also verified by an index calculation. 
We also discuss a factorisation of this result and compare it with analogous results for $S^7$.

\end{abstract}

\eject
\normalsize

\tableofcontents

\section{Introduction}

Since Pestun \cite{Pestun:2007rz} calculated the exact partition function for $\N = 2$ supersymmetric gauge theories on $S^4$, supersymmetric localisation of gauge theories has attracted much attention. A recent and thorough review is given in \cite{Pestun:2016zxk}. The methods in \cite{Pestun:2007rz} have been generalised to other supersymmetric theories in various dimensions and geometries, see e.g. \cite{Pestun:2016jze} for an overview. The methods and results fall roughly into two categories depending on if the dimension is even or odd. Since we are discussing 7D in this paper, let us restrict the discussion to the odd dimensional case.

In 3D there is a rich class of supersymmetric theories and e.g. localisation of 3D $\N = 2$ theories on compact manifolds is a well studied subject, see \cite{willett2017loc3dRev} for a review. Rigid supersymmetric theories with four or fewer supercharges in dimension four or less are classified, but in higher dimensions (and also for eight supercharges in 4D) the classification remains open. Although the supergravity approach initiated by Festuccia-Seiberg \cite{Festuccia:2011ws} could in principle provide such a classification, it becomes increasingly complicated in higher dimensions.

Apart from performing the localisation there is also the issue of carrying through the calculation and writing the answer in a closed form (e.g. in terms of special functions). To be able to do this  one typically needs some extra symmetry of the manifold. Usually this is in the form of a  toric action, and we will discuss this point for 7D 3-Sasakian manifolds later on.

In 5D there are infinite families of toric Sasaki-Einstein manifolds for which one can derive the perturbative partition function for SYM coupled to hypermultiplets, see \cite{Qiu:2016rev} for a review. The perturbative answer is given by a matrix model involving generalised triple sine functions governed by the toric geomerty. Finding the full partition function becomes trickier in higher dimensions due to the localisation locus equations becoming more complicated. However, using factorisation results for the perturbative answer one can conjecture that the full partition function also obey the same factorisation and write the full partition function in terms of Nekrasov partition functions \cite{Qiu:2014oqa}. 

Moving up to 7D the SYM theory is unique. It is the maximally supersymmetric Yang-Mills, and thus it is only the geometry of the underlying 7D manifold that governs the theory. The case of $S^7$ was studied in \cite{Minahan:2015jta} and this was then generalised to 7D toric Sasaki-Einstein manifolds in \cite{Polydorou2017}. As in the 5D case, the perturbative partition function can be found in terms of toric data and is given by a matrix model involving generalised quadruple sine functions. One can speculate about the full partition function based on factorisation properties similar to those in 5D, however the basis for such a conjecture is much weaker, see \cite{Polydorou2017} for a discussion. Another issue is that maximally SYM is non-renormalisable and hence requires UV-completion. However, the localisation procedure is well-defined and produces finite results for observables and it is expected that these can be interpreted in the UV-completed theory. For the maximally supersymmetric theory in 5D it is believed that the UV-completion is the elusive 6D (2,0) SCFT and one can indeed find agreement for some specific results, see e.g. \cite{Kallen:2012zn}. 
For 7D we do not know what the corresponding UV-completion is and although we can perform well-defined localisation calculations it is unclear what to match them with. A speculation is that it has something to do with `little' $m$-theory \cite{Losev:1997hx}.

Let us briefly comment on why we do not go higher than 7 dimensions. For spheres, $S^d$, there is simply no suitable superalgebra for $d>7$, a fact related to Nahm's classification of superconformal theories \cite{Nahm1978}. The approach of placing on-shell maximally supersymmetric Yang-Mills on curved manifolds by dimensional reduction from flat space also limits the dimension to be no more than seven, see \cite{Blau}.

The number of supersymmetries of the theory is intimately related to geometry and is determined by the number of positive Killing spinors the manifold admits. 
In 5D the manifolds admitting Killing spinors are Sasaki-Einstein manifolds (2 Killing spinors) and the five-sphere (8 Killing spinors). In 7D there are more possibilities: proper $G_2$ manifolds (1 Killing spinor), Sasaki-Einstein manifolds (2 Killing spinors), 3-Sasakian manifolds (3 Killing spinors) and $S^7$ (16 Killing spinors). One thing that makes 7D special is thus the possibility of $G_2$ or 3-Sasakian structures\footnote{$G_2$-structures are unique to dimension 7. 3-Sasakian structures can exist in dimensions $4n-1$, however in dimension three all Sasaki-Einstein manifolds are also 3-Sasakian. Thus dimension 7 is the smallest dimension for which these two notions are distinct.}. 
The seven-sphere was considered in \cite{Minahan:2015jta}, toric Sasaki-Einstein manifolds in \cite{Polydorou2017}, and in this paper we will discuss a 3-Sasakian manifold. Although one can define supersymmetric Yang-Mills on proper $G_2$-manifolds, it is unclear how to take it off-shell, see \cite{Polydorou2017} for a discussion, and we leave it for future work.

For 3-Sasakian manifolds we have three Killing spinors and the action is invariant under three supersymmetry transformations.
One would expect this extra supersymmetry to be manifest somehow in the final result of the localisation calculation. In \cite{Polydorou2017} some first steps were made towards understanding the role of the 3-Sasakian structure for localisation on $S^7$. In particular, different factorisations of the partition function were discussed involving $SU(2)$-structures. This was based on viewing $S^7$ as an $SU(2)$-fibration over $S^4$ via the (quaternionic) Hopf fibration. 

In this paper we continue the study of 7D SYM on 3-Sasakian manifolds by computing the perturbative partition function for an explicit example outside the scope of \cite{Polydorou2017}. The reason we consider an example is that we do not yet have the tools to analyse the general case.

The reason our example is interesting is firstly because it is 3-Sasakian and thus admits three supersymmetries. Secondly, unlike $S^7$, it is not toric. The lack of an effective hamiltonian $T^4$-action on the metric cone of the manifold means that the techniques of \cite{Minahan:2015jta,Polydorou2017} cannot be applied. However, the hyperk\"ahler metric cone of our manifold has an effective hyperhamiltonian $T^2$-action which makes it a `hypertoric  variety'. This hypertoric symmetry is exploited in our calculations and we believe the techniques presented here can also be applied to other 3-Sasakian manifolds whose hyperk\"ahler cones are hypertoric.

This paper is organised as follows: in section \ref{sec:7DSYM} we review 7D SYM and the localistaion for toric Sasaki-Einstein manifolds, including $S^7$. In section \ref{sec:GeoSetting} we give a quick review of 3-Sasakian geometry and then introduce the specific 7D 3-Sasakian manifold for which we will perform the localisation calculation. In section \ref{sec:HoloFunCalc} we discuss hypertoric geometry and use it to find the perturbative partition function via a count of holomorphic functions. In section \ref{sec:factor} we discuss a factorisation of the result in terms of $SO(3)$ data. We finish with a summary and directions for future work in section \ref{sec:summary}. Further details of calculations and conventions are provided in the appendices.

\section{7D supersymmetric Yang-Mills} \label{sec:7DSYM}

In \cite{Minahan:2015jta} supersymmetric Yang-Mills on spheres was studied. Starting from maximally supersymmetric Yang-Mills on 10-dimensional Lorentzian flat space a supersymmetric action on $S^7$ was obtained by dimensional reduction and deformation:
\begin{align} \label{7Daction}
S_{7D}= \frac{1}{g_{7D}^2} \int d^{7}x \sqrt{-g} \Tr \Big( 
\halfs F^{MN}F_{MN} 
- \Psi \G^M D_M \Psi 
+8 \phi^A \phi_A
+\tfrac32 \Psi \Lambda \Psi
-2 [\phi^A,\phi^B]\phi^C \varepsilon_{ABC}
\Big)\, .
\end{align}
The above formula is in 10-dimensional notation where the indices $M,N=0,\dots,9$ while $A,B,C$ run over the compactified dimensions $A,B,C=0,8,9$. Here $F_{MN}$ is the field strength of the gauge field $A_M$, $\Psi$ a Majorana-Weyl fermion transforming in the adjoint representation of the gauge group, and $\phi_A$ the scalars that come from the dimensional reduction. The $\Gamma^M$ are 10-dimensional Dirac matrices and $\Lambda$ a product of three such matrices. The $\varepsilon_{ABC}$ denotes the anti-symmetric symbol in the compactified dimensions. A trace is taken over the colour indices (which are omitted in the notation). We refer the reader to \cite{Minahan:2015jta} or \cite{Polydorou2017} for more details. 

The supersymmetry transformations for this action are 
\begin{align} 
\delta_\eps  A_M &= \eps \G_M \Psi \notag \, ,\\
\delta_\eps \Psi &= \tfrac{1}{2} F_{MN}\G^{MN}\eps + \tfrac{8}{7} \G^{\mu B} \phi_B \nabla_\mu \eps\, , \label{susytranson}
\end{align}
where $\mu =1,\dots, 7$. This transformation relies on the existence of a 10-dimensional Majorana-Weyl spinor $\eps$ satisfying a generalised Killing spinor equation.
In \cite{Minahan:2015jta} such a spinor was constructed using conformal Killing spinors on $S^7$ and in \cite{Polydorou2017} this was generalised to other compact 7D manifolds admitting positive Killing spinors. Such manifolds fall into four categories: proper $G_2$ manifolds (1 Killing spinor), Sasaki-Einstein manifolds (2 Killing spinors), 3-Sasakian manifolds (3 Killing spinors) and $S^7$ (16 Killing spinors).

In order to perform the localisation arguments the supersymmetry needs to be taken off-shell and in \cite{Minahan:2015jta, Polydorou2017} this was done by introducing pure spinors and auxiliary fields. It was shown how the fields could be mapped to differential forms and how the supersymmetry could be written as a cohomological complex for Sasaki-Einstein manifolds (which includes 3-Sasakian and $S^7$ as subcases). It was argued in \cite{Polydorou2017} that the vector field $R^\mu = \eps \Gamma^\mu \eps$ serves as the Reeb vector field for the Sasaki-Einstein case but is identically zero for proper $G_2$-manifolds, thus the latter case is excluded from the analysis.

After gauge-fixing the following cohomological complex was found \cite{Polydorou2017}:
\begin{align}
(A,\psi),(\Phi,\eta), (\Upsilon,H), (c,\phi_0), (\bar{c},b), (b_0,c_0), (\bar{a}_0,\bar{c}_0)  \,,
\end{align}
where each $(X,X')$-pair satisfies
\begin{align}
Q X &= X'\, , \nn\\
Q X'&=  (-\L_R + iG_{a_0}) X \, .
\end{align}
Here $Q=\delta_\eps + \delta_{BRST}$ is the supersymmetry combined with the standard BRST transformation.
Let us briefly sketch how the above complex was found, details can be found in \cite{Polydorou2017}.
The gauge fields $A_\mu$ were mapped to the connection $A$, the bosons $\phi_8,\phi_9$ to the 3-form $\Phi$, and the bosonic auxiliary fields mapped to the 2-form $H$. The fermion field $\Psi$ was mapped to the 1-form $\psi$, the 2-form $\Upsilon$ and the 3-form $\eta$. The gauge fixing introduced the ghosts $c,\bar{c}$, the Lagrange multiplier $b$ and the zero modes $a_0,\bar{a}_0,b_0, c_0, \bar{c}_0$, with the $a$'s and $b$'s bosonic and the $c$'s fermionic. Note that $Q^2 = -\L_R + iG_{a_0}$, i.e. it squares to a Lie derivative and a gauge transformation.

The simplest solution to the fixed point locus equations in \cite{Polydorou2017} is $A=0$, $\Phi=0$ and $\phi_0=a_0=constant = \sigma$ and the full perturbative partition function is obtained as the one-loop approximation:
\begin{equation} \label{partfun1}
\int\limits_g d\sigma e^{- \frac{24}{g^2_7} V_7 \Tr(\sigma^2)} \frac{ \sqrt{det_{\Omega_H^{(2,0)}}(Q^2)det_{\Omega_H^{(0,2)}}(Q^2) det_{\Omega^{0}}(Q^2)}\sqrt{det_{\Omega^{0}}(Q^2)}
\sqrt{det_{\Omega^{0}}(Q^2)} }{
\sqrt{det_{\Omega^1}(Q^2)}
\sqrt{det_{\Omega_H^{(3,0)}}(Q^2)det_{\Omega_H^{(0,3)}}(Q^2)}
\sqrt{det_{H^0}(Q^2)}
\sqrt{det_{H^0}(Q^2)}} \, .
\end{equation}
The terms in the exponent come from evaluating the action at the fixed point (here $V_7$ denotes the volume of the 7-dimensional manifold). The determinant factors in the numerator come from integrating over the fermions $(\Upsilon, c, \bar{c})$ and the denominator from the bosons $(A,\Phi,b_0,\bar{a}_0)$. Note that the forms are Lie algebra valued. It was further argued in \cite{Polydorou2017} that this simplifies to
\begin{align} \label{partfun2}
Z^{pert}
&=\int\limits_g d\sigma e^{-\frac{24}{g^2_7} V_7 \Tr(\sigma^2)} \frac{det_{\Omega^{0}}(Q^2) det_{\Omega_H^{(0,2)}}(Q^2)}{det_{\Omega_H^{(0,1)}}(Q^2)det_{\Omega_H^{(0,3)}}(Q^2)} \\
&= \int\limits_g d\sigma e^{-\frac{24}{g^2_{7D}} V_7 \Tr(\sigma^2)}  det'_{adj}\, sdet_{\Omega_H^{(0,\bullet)}}(-\L_R + iG_{\sigma})\, ,
\end{align}
where the determinant over the adjoint representation of the Lie algebra was written out explicitly in the last step.

This answer is structurally very similar to that in 5D, see e.g. \cite{Qiu:2016rev} for a review, and the task reduces to computing the superdeterminant
\begin{equation}
sdet_{\Omega_H^{(0,\bullet)}}(-\L_R + x) \label{eq:sdet} \, ,
\end{equation}
where $\L_R$ denotes the Lie derivative along the Reeb vector field $R$ and $\Omega_H^{(0,\bullet)}$ denotes horizontal $(0,p)$-forms. Let us take a moment to explain these geometrical structures and how to evaluate this superdeterminant.

Let $X$ denote the 7D Sasaki-Einstein manifold. We define its metric cone $C(X)$ as
\begin{equation}
C(X) = X \times \R^+\, ,
\end{equation}
with metric 
\begin{equation}
ds_{C(X)}^2 = d r^2 + r^2 ds_{X}^2\, .
\end{equation}
Here $r$ is the coordinate on $\R^+$ and $ds_{X}^2$ is the metric on $X$.
Recall that $X$ is Sasaki-Einstein if its metric cone is Calabi-Yau. The Calabi-Yau structure on $C(X)$ induces several nice structures on $X$. For example we get a contact form $\k$ and its associated Reeb vector field $R$, satisfying $\iota_R \k = 1$ and $\iota_R d\k =0$. The contact structure gives rise to a horizontal space ($\ker \k$) and we can talk about horizontal differential forms $\Omega_H$. Moreover, we have an almost complex structure on the horizontal space which allows the definition of Dolbeault operators on the horizontal forms. The differential operator
\begin{equation}
\db_H: \Omega_H^{(p,q)} \rightarrow \Omega_H^{(p,q+1)}
\end{equation}
gives rise to the Kohn-Rossi cohomology groups $H_{KR}^{(p,q)}$. One can show (see e.g. \cite{Schmude:2014lfa}) that $\L_R$ commutes with $\db_H$ and \eqref{eq:sdet} thus reduces to a superdeterminant over the cohomology groups $H_{KR}^{(0,\bullet)}$. There are two main ways of proceeding with the superdeterminant calculation from here.

The first approach is to think of the superdeterminant as the index of the operator $\db_H$. This is a transversally elliptic operator and its index can be computed using methods in Atiyah's book \cite{AtiyahElliptic}. This was the original approach of Qiu and Zabzine for the 5D calculation in \cite{Qiu:2013pta}.

The second approach is due to Schmude \cite{Schmude:2014lfa} and calculates the superdeterminant in terms of holomorphic functions on the cone $C(X)$. Let us sketch this argument in 7D. The Calabi-Yau structure on $C(X)$ gives rise to a nowhere vanishing $(3,0)$-form on the horizontal space of $X$. This form provides a pairing between $(0,0)$-forms and $(0,3)$-forms which is also a pairing in cohomology, i.e. $H_{KR}^{(0,0)} \iso H_{KR}^{(0,3)}$. However, the $\L_R$ eigenvalues differ by an overall minus sign and a shift corresponding to the eigenvalue of the $(3,0)$-form used in the pairing. One can also obtain a pairing $H_{KR}^{(0,1)} \iso H_{KR}^{(0,2)}$ but by simply-connectedness the former is zero. What remains of the superdeterminant calculation is thus to find $H_{KR}^{(0,0)}$ (and the shift in eigenvalues resulting from the pairing via the $(3,0)$-form). But 
$H_{KR}^{(0,0)} \iso H^0(\mathcal{O}_{C(X)})$, so this can be found by counting the holomorphic functions on the Calabi-Yau cone $C(X)$. For \emph{toric} Sasaki-Einstein manifolds this count has a nice combinatorial description in terms of toric data and the superdeterminant can be written in terms of generalised multiple sine functions.

For 7D toric Sasaki-Einstein manifolds we get
\begin{equation}
sdet_{\Omega_H^{(0,\bullet)}}(-\L_R + x) \sim S_4^{C_\mu(X)}(x|\vec{R}) \label{eq:sdetS4} \, ,
\end{equation}
where $S_4^{C_\mu(X)}$ is the generalised quadruple sine function associated to the moment map cone $C_\mu(X)$ of the torus action, see \cite{Polydorou2017} for details.

The final result for the perturbative part of the partition function is then \cite{Polydorou2017}
\begin{align}
Z^{\text{pert}}
&= \int\limits_t d \sigma \,  e^{-\frac{24}{g^2_7} V_7 \Tr(\sigma^2)} \prod_\beta S_4^{C_\mu(X)}( i \bra \sigma, \beta \ket |\vec{R})\, , \label{pertparfun}
\end{align}
where $\beta$ are the roots of the Lie algebra $g$ 
and the Weyl integration formula was used to reduce the integral to the Cartan subalgebra $t$.

For the special case of $S^7$, the generalised quadruple sine reduces reduces to the ordinary quadruple sine, which is described by the infinite product expression
\begin{equation}
S_4 (x|\w_1,\w_2,\w_3,\w_4) = \frac{\prod\limits_{n_1,n_2,n_3,n_4 \geq 0}\left( n_1\w_1+n_2\w_2+n_3\w_3+n_4\w_4 + x \right) } {\prod\limits_{n_1,n_2,n_3,n_4 \geq 1}\left( n_1\w_1+n_2\w_2+n_3\w_3+n_4\w_4 - x \right)} \, .
\label{eq:S4Defn}
\end{equation}

In this paper we will calculate the superdeterminant \eqref{eq:sdet}, and hence the perturbative partition function, for an explicit 7D example that lies outside the scope of \cite{Polydorou2017}. The manifold we will consider is a 7D 3-Sasakian manifold (and hence Sasaki-Einstein) but it is not toric. The arguments of \cite{Polydorou2017} can thus be used, but only up to the final evaluation of the superdeterminant.
For our new example, we will evaluate the superdeterminant in two ways. 
First we will again count holomorphic functions on the cone $C(X)$, which in the 3-Sasakian case is hyperk\"ahler. This will use that $C(X)$ is `hypertoric' - a hyperk\"ahler analogue of toric. We will then also verify our result via an index calculation. 
Let us now proceed to describe the manifold for which we will do this calculation.

\section{Geometrical setting} \label{sec:GeoSetting}

In this section we will give the geometrical description of the 7D 3-Sasakian manifold $\S$ that will be our main example. We have not found a standard name for this manifold in the literature but since it is the total space of a principal $SO(3)$-bundle over $\CP^2$ it might be referred to as the `Konishi bundle over $\CP^2$'. One can also characterise this manifold as the 3-Sasakian manifold whose metric cone is the `Swann bundle over $Gr_2(\C^3)$', or whose `associated twistor space' is the complete flag $F_{1,2}(\C^3)$.

\subsection{3-Sasakian manifolds and related structures}
Let us start by recalling some facts about 3-Sasakian geometry. We refer the reader to \cite{Boyer1998} for a thorough introduction. A 3-Sasakian manifold is perhaps most easily defined as a manifold $X$ whose metric cone $C(X)$ is hyperk\"ahler. This means that there are three almost complex structures $I, J, K$ on $C(X)$ satisfying the quaternionic relations
\begin{equation}
I^2 =J^2=K^2=IJK=-1.
\end{equation}
Thinking of $X$ as the base of the cone at $r=1$, we can obtain three vector fields on $X$ via
\begin{equation}
R_a = I_a(r \partial_r)|_{r=1}, \quad I_a = I,J,K \, , 
\end{equation}
and three one-forms via
\begin{equation}
\k_a(Y)=g(R_a,Y).
\end{equation}
These give the three Reeb vectors $R_a$ and contact forms $\k_a$ of the 3-Sasakian structure on $X$. They satisfy the relations
\begin{align}
\iota_{R_a} \k_b &= \delta_{ab}, \\
[R_a,R_b]  &= \epsilon_{abc} R_c. \label{eq:ReebSU2Alg}
\end{align}
Apart from the cone, every 3-Sasakian manifold also come together with two other geometrical objects: the quaternionic K\"ahler orbifold $\O$ and the twistor space $\mathcal{Z}$. These objects and their relations are summarised in the `fundamental diagram' \cite{Boyer1998}:
\begin{equation}
\begin{tikzcd}[column sep=small]
& \arrow[dl] C(X) \arrow{dd} \arrow[dr]& \\
\mathcal{Z} \arrow[dr]  &  & \arrow{ll} X \arrow[dl] \\
& \O & 
\end{tikzcd} \label{FundDiag}
\end{equation}
Let us briefly explain the relations in this diagram (again we refer to \cite{Boyer1998} for details). The arrows from $X$ to $\mathcal{Z}$ and $\mathcal{O}$ correspond to the one and three-dimensional 3-Sasakian foliations respectively. The arrow from $\mathcal{Z}$ to $\O$ corresponds to the twistor map. The arrow from $C(X)$ to $\O$ corresponds to the so called Swann bundle \cite{Swann1991} which can be seen as a hyperk\"ahler generalisation of the twistor space. 
One can go from the Swann bundle to the twistor space $\mathcal{Z}$ by fixing a complex strucutre and projectivising \cite{Swann1991}. Finally, the arrow from $C(X)$ to $X$ denotes the cone relation

For 7-dimensional $X$, the most familiar example of this picture is probably
\begin{equation}
\begin{tikzcd}[column sep=small]
& \arrow[dl] \C^4\setminus 0  \arrow{dd} \arrow[dr]& \\
\CP^3 \arrow[dr]  &  & \arrow{ll} S^7 \arrow[dl] \\
& \mathbb{H}P^1 \iso S^4 & 
\end{tikzcd} \label{S7FundDiag}
\end{equation}
where the 1- and 3-foliations of $S^7$ correspond to the complex and quaternionic Hopf fibrations respectively.

Another 7D example, which is the one we will focus on here, is
\begin{equation}
\begin{tikzcd}[column sep=small]
& \arrow[dl] C(\S) = \mathcal{U}(Gr_{2}(\C^3))  \arrow{dd} \arrow[dr]& \\
F_{1,2} (\C^3) \arrow[dr]  &  & \arrow{ll} \S \arrow[dl] \\
& Gr_{2}(\C^3) \iso \CP^2 & 
\end{tikzcd}  \label{SwannFundDiag}
\end{equation}
Here $Gr_2(\C^3)$ denotes the Grassmannian of of 2-planes in $\C^3$ and $\mathcal{U}(Gr_{2}(\C^3))$ denotes its Swann bundle. Note that $Gr_2(\C^3)$ is isomorphic to $\CP^2$ (but with the non-canonical orientation). The twistor space is $F_{1,2}(\C^3)$,  i.e. the complete flag manifold in $\C^3$. The 7D 3-Sasakian manifold $\S$ in this picture will be the main object of study in this paper. 

Let us also remark that another viewpoint for the diagram \eqref{FundDiag} is to take the quaternionic K\"ahler orbifold $\O$ of positive scalar curvature as the starting point. 
Let us restrict to the case when $\O$ is a {smooth manifold}, as is the case in our examples above\footnote{In fact these are the \emph{only} such examples in dimension 4 \cite{hitchin1981kahlerian}. However, for our purposes we 
do not need $\mathcal{O}$ to be smooth, only $X$, so we are not restricted to just these two cases.}. There are a few ways to get the 3-Sasakian manifold $X$ in \eqref{FundDiag} from $\O$.
One could first construct the twistor space \cite{salamon1982quaternionic} of $\O$ and then get to $X$ by using an inversion theorem from \cite{boyer1997twistor}. Alternatively, one could construct the Swann bundle \cite{Swann1991} over $\O$ and then get the 3-Sasakian manifold $X$ as a hypersurface in this space \cite{boyer1993quaternionic}. Finally, one could take the Konishi bundle \cite{konishi1975} over $\O$ to obtain the 3-Sasakian manifold as an $SU(2)$ or $SO(3)$-bundle. These constructions can also be generalised to the case when $\O$ is an {orbifold} \cite{boyer1993quaternionic,Boyer1998}.

\subsection{The 3-Sasakian manifold $\S$} \label{sec:TheManifoldS}

Let us now focus on the 7D 3-Sasakian manifold $\S$ whose fundamental diagram is given by \eqref{SwannFundDiag}. It is on this manifold we will place our SYM theory and compute the perturbative partition function. First we will obtain this manifold as a hypersurface in the Swann bundle $\mathcal{U}(Gr_{2}(\C^3))$ and then explain how it is an $SO(3)$ bundle over $\CP^2$.

Following \cite{Swann1991} we will realise the Swann bundle $\mathcal{U}(Gr_{2}(\C^3))$ as a hyperk\"ahler quotient\footnote{The construction of this space as a bundle can also be found in \cite{Swann1991}.}. It is convenient to formulate this in terms of quaternions, see Appendix \ref{AppQuaternions} for our conventions about the flat quaternionic space $\H$.

We start in $\H^3$ with coordinates $\vq$. We will map this to $V \times V^*$, where $V=\C^3$ and $V^*$ its dual, with coordinates $\vz, \vw$ via $\vq = \vz + j \vw$. The complex structures, symplectic forms, moment maps etc will then just be three copies of those described for the quaternions in Appendix \ref{AppQuaternions}.

We take the hyperk\"ahler quotient 
\begin{equation}
V \times V^* /// U(1)_Q \, ,
\end{equation}
where $U(1)_Q$ is generated by left multiplication by $i$ on $\vq$. The charge vector is given by $Q = [1, 1, 1; -1, -1, -1]$. The real and complex moment maps are
\begin{align}
\mu_Q^{\R} &= -\frac{1}{2}(|\vz|^2 - |\vw|^2) \, , \label{eq:mRSwann}\\
\mu_Q^{\C} &= i \vz \cdot \vw \,. \label{eq:mCSwann}
\end{align}
Taking the zero level-sets of the real and complex moment maps (and quotienting by $U(1)_Q$) gives us the Swann bundle $C(\S)= \U (Gr_2(\C^3))$. 

Note that if we instead take a non-zero level-set of the real moment map we would end up with $T^* \CP^2$ with its Calabi metric \cite{calabi1979metriques}.

We can characterize $C(\S)= \U (Gr_2(\C^3))$ as $(\vz, \vw) \in V \times V^*$ subject to the relations
\begin{align}
|\vz|^2 &= |\vw|^2 \, ,  \label{eq:CS1}\\
\vz \cdot \vw &= 0 \, , \\
(\vz, \vw) &\sim (e^{i\theta} \vz, e^{-i \theta} \vw) \, .
\end{align}
The cone structure of $C(\S)$ is seen from \eqref{eq:CS1} and we obtain $\S$ as a hypersurface (base of the cone) by setting $r = |\vz| = |\vw| = 1$.

In $\H^3$ we also have three complex structures, $I,J,K$, generated by \emph{right} multiplication by $i, j, k$. These are invariant under \emph{left} multiplication by $i$ and thus they descend to the quotient. In terms of the coordinates $\vz, \vw$ we can write them as
\begin{align}
I &= i d z^i \otimes \partial_{z^i} + i d{w^i} \otimes \partial_{w^i} + c.c.  \label{eq:acsI} \\
J &= - d{\bar{w}^i} \otimes \partial_{z^i} +  d \bar{z}^i \otimes \partial_{w^i} + c.c. \\
K &= i d{\bar{w}^i} \otimes \partial_{z^i} - i d \bar{z}^i \otimes \partial_{w^i} + c.c. 
\end{align}
From this we can find the Reeb vectors on $\S$ as follows. The cone coordinate $r$ can be written as 
\begin{align}
r^2 &= \frac{1}{2} |\vq|^2 = \frac{1}{2} \left( |\vz|^2 + |\vw|^2 \right) \, ,
\end{align}
and thus the homothetic vector field is given by
\begin{equation}
r \partial_r =  q^i \frac{\partial}{\partial {q^i}} = z^i \partial_{z^i} + w^i \partial_{w^i} + c.c.
\end{equation}
From this vector field we obtain the three Reebs via
\begin{align}
R_1 &= I(r \partial_r) |_{r=1} = i z^i \partial_{z^i} + iw^i \partial_{w^i} + c.c. \label{eq:Reeb1} \, ,\\
R_2 &= J(r \partial_r) |_{r=1} = -\bar{w}^i \partial_{z^i} + \bar{z}^i \partial_{w^i} + c.c. \label{eq:Reeb2} \, ,\\
R_3 &= K(r \partial_r) |_{r=1} = i\bar{w}^i \partial_{z^i} -i \bar{z}^i \partial_{w^i} + c.c.  \label{eq:Reeb3} \, .
\end{align}
The three Reebs are generated by right action of $i,j,k$, and $U(1)_Q$ is the left action of $i$. Since left and right actions commute, the $R_a$ and $U(1)_Q$ commute. The Reebs also preserve the zero sets of the moment maps and hence we get three vector fields $R_a$ on $\S$ satisfying the quaternionic algebra i.e. the $SU(2)$ algebra of \eqref{eq:ReebSU2Alg}.

Let us now explain how $\S$ is an $SO(3)$ bundle over $\CP^2$. Viewing $\S$ as the base of the cone we have $|\vz|^2 = |\vw|^2 = 1$ as well as $\vz \cdot \vw = 0$. Thus $|\vz \times \vw^*|^2 = |\vz|^2 |\vw|^2 = 1$ and so $\vz \times \vw^*$ parametrises $S^5$ (here $w^*$ denotes complex conjugation). However, $U(1)_Q$ acts diagonally on $\vz \times \vw^*$ and so the reduction by $U(1)_Q$ gives us $\CP^2$, parametrised by $[\vz \times \vw^*]$. 
One might expect the bundle $\S \rightarrow \CP^2$ to have fibre $SU(2)$ (as is the case for the quaternionic Hopf fibration of $S^7$, c.f. \eqref{S7FundDiag}), but there is an important subtlety  here. Consider the Reeb vector $R_1$ and its corresponding $U(1)$ action with charge $[1,1,1;1,1,1]$ and angle variable $\theta$. When $\theta = \pi$ the effect can be undone by $U(1)_Q$:
\begin{equation}
( e^{i \pi} \vz, e^{i\pi}\vw ) \sim ( e^{i \pi} e^{i \pi} \vz, e^{-i \pi} e^{i\pi}\vw ) = ( e^{2 \pi i} \vz, e^{0}\vw ) =  (\vz,\vw) \, . \label{eq:SO3Fibre}
\end{equation} 
Hence the fibre is not $SU(2)$ but $SU(2)/\Z_2 \iso SO(3)$.

Finally, we note that the Reebs \eqref{eq:Reeb1}-\eqref{eq:Reeb3} act trivially on the base $\vz \times \vw^*$, for example
\begin{align}
R_1(\vz \times \vw^*) &= i\vz \times \vw^* + \vz \times (-i \vw^*) = 0 \, .  
\end{align}
One can check that $R_a \perp U(1)_Q$ and thus compute their norms on the flat space and get $|R_a|^2=2$. We thus have a locally free action of $SO(3)$ on $\S$. This action is also free.

In order to find the perturbative partition function of our SYM theory we need to compute the superdeterminant \eqref{eq:sdet}. To do this calculation for the manifold $\S$ presented above, we will use that fact that the cone $C(\S)$ is a hypertoric variety. We now turn to discussing this point.

\subsection{Hypertoric geometry}

Recall that a 7D manifold $X$ is Sasaki-Einstein if its metric cone is Calabi-Yau. Moreover, we say that it is \emph{toric} if $C(X)$ admits an effective hamiltonian action of the 4-torus $T^4$. We also require that the Reeb vector field can be written as a linear combination of the four $U(1)$'s of the torus action.

For a 3-Sasakian $X$, its metric cone $C(X)$ is hyperk\"ahler and thus has three complex structures, symplectic forms etc. The action of a compact Lie group on $C(X)$ is said to be \emph{hyperhamiltonian} if it is hamiltonian with respect to each three of the symplectic forms.
The 8D hyperk\"ahler manifold $C(X)$ is called \emph{hypertoric} if it admits an effective hyperhamiltonian action of the 2-torus $T^2$.

Hypertoric\footnote{Note that the terms `hypertoric' and `toric hyperk\"ahler' are used interchangably in the literature.} geometry, first introduced by Bielawski and Dancer \cite{Bielawski2000}, is by now a well-studied area in mathematics, see e.g. \cite{Proudfoot2008survey, Konno2008survey} for reviews. It is also of interest to physicists since, for example, the Higgs and Coulomb branches of abelian 3D $\N=4$ gauge theories  \cite{deBoer1997mirror} are hypertoric varieties.

Note that while 3-Sasakian manifolds are also Sasaki-Einstein, hypertoric manifolds are \emph{not} necessarily  toric. So while the reasoning based on the Sasaki-Einstein structure of $X$ in \cite{Polydorou2017} also applies to the 3-Sasakian case, the arguments for \emph{toric} $C(X)$ do not apply to the hypertoric case. Note however that $S^7$, whose metric cone is $\C^4$, is both toric \emph{and} hypertoric\footnote{More generally, Hausel and Sturmfels \cite{Hausel2002} have shown that the only manifolds that are both toric and hypertoric are ALE spaces of type $A_n$ or products of such spaces.}.

Let us focus on hypertoric manifolds that arise as hyperk\"ahler quotients of flat quaternionic space. (By a theorem of Bielawski \cite{Bielawski1999} all hypertoric manifolds with Euclidean volume growth can be obtained in this way.) Following \cite{Bielawski2000} we now describe this construction.

Let $\H^n$ be the quaternionic vector space with the standard hyperk\"ahler structure given by left multiplication by $i, j, k$, see Appendix \ref{AppQuaternions}. As before, we identify $\vq\in \H^n$ with $(\vz,\vw) \in V \times V^*$, where $V=\C^n$, via $\vq = \vz + j \vw$. 
The real torus $T^n = \{t=(t_1,\dots,t_n) \in \C^n | |t_i|=1 \}$ acts on $\H^n$ by left diagonal multiplication $t(z,w) = (t\vz,t^{-1} \vw)$ and it preserves the hyperk\"ahler structure.
The corresponding moment maps are
\begin{align}
\mu_\R(\vz,\vw) &= -\frac12 \sum_{k=1}^n (|z_k|^2-|w_k|^2)e_k + c_1 \label{eq:mRstdTorus} \, , \\
\mu_\C(\vz,\vw) &=  i\sum_{k=1}^n (z_k w_k)e_k + c_2 + ic_3 \, , \label{eq:mCstdTorus} 
\end{align}
where $e_k$ denotes the standard basis of $\R^n \iso \mathfrak{g}^*$ and $c_i$ are arbitrary central elements which we think of as constant vectors in $\R^n$.

Now consider a subtorus $K$ of $T^n$ with Lie algebra $\mathfrak{k}$, and call $T^d = T^n/K$. We represent it by a collection of $n$ vectors $\{u_i\}$ in $\R^d$. We take these to be non-zero, integer, primitive 
vectors that generate $\R^d$. These define a map $\beta: \mathfrak{t}^n \iso \R^n \rightarrow \mathfrak{t}^d \iso \R^d$ via $e_i \mapsto u_i$.
This gives rise to an exact sequence 
\begin{equation}
\begin{tikzcd}[column sep=small]
0 \arrow{r} &  \mathfrak{k} \arrow{r}{\iota} & \mathfrak{t}^n \arrow{r}{\beta} & \mathfrak{t}^d \arrow{r} & 0 
\end{tikzcd}  \, ,\label{eq:exactseq1}
\end{equation}
and its dual
\begin{equation}
\begin{tikzcd}[column sep=small]
0 \arrow{r} &  (\mathfrak{t}^d)^* \arrow{r}{\beta^*} & (\mathfrak{t}^n)^* \arrow{r}{\iota^*} & \mathfrak{k}^* \arrow{r} & 0 
\end{tikzcd} \, .
\end{equation}
The moment maps for the subtorus action are then given by 
\begin{align}
\mu_\R(z,w) &= -\frac12 \sum_{k=1}^n (|z_k|^2-|w_k|^2)a_k + c_1 \, ,\label{eq:mRsubTorus}\\
\mu_\C(z,w) &=  i\sum_{k=1}^n (z_k w_k)a_k + c_2 + ic_3 \, , \label{eq:mCsubTorus}
\end{align}
where $a_k = \iota^*(e_k)$.
The constants $c_i$ are of the form $c_i = \sum_{k=1}^n \lambda_k^i a_k$. We will use the compact notation $\lambda_k = (\lambda_k^1,\lambda_k^2,\lambda_k^3)$.
Following \cite{Bielawski2000} we denote the hyperk\"ahler quotient $\mu^{-1}(0)/K$ corresponding to the data $\vec{u} = (u_1,\dots,u_n)$, $\vec{\lambda} = (\lambda_1,\dots,\lambda_n)$ by $M(\vec{u},\vec{\lambda})$. The action of $T^d$ on $M(\vec{u},\vec{\lambda})$ preserves the hyperk\"ahler structure and $M(\vec{u},\vec{\lambda})$ is a \emph{hypertoric variety}.

From the data $(\vec{u},\vec{\lambda})$ we define \emph{hyperplanes} in $\R^d$ by
\begin{equation} \label{hypPlaneDef}
H^i_k = \{ y\in \R^d | y \cdot u_k = \lambda_k^i \}
\end{equation}
and use the compact notation $H_k = H_k^1 \times H_k^2 \times H_k^3 \subset \R^{3d}$.

Recall that for toric manifolds the geometry is determined by the Delzant polytope which can be described in terms of a set of hyperplanes with normal vectors $\{v_i\}$. Similarly, in the hypertoric case the geometry is determined by the arrangement of the hyperplanes $H_k$, which is encoded by their normal vectors $\{u_i \}$ and the $\lambda$'s.

Many of the statments from toric geometry relating geometrical properties to polytope data have hypertoric analogues in terms of hyperplanes. For example, we have the following smoothness condition for hypertoric varieties \cite{Bielawski2000}:
Let $\vec{u},\vec{\lambda}$ be such that the $H_k$ are distinct. Then $M(\vec{u},\vec{\lambda})$ is smooth iff whenever a set of hyperplanes $\{ H_l \}$ intersect in $\R^{3d}$, their corresponding normal vectors $\{ u_l \}$ can be completed into a $\Z$-basis of $\Z^d$. (In particular, this means that any $d+1$ hyperplanes must have empty intersection.)
This is equivalent to the `good cone condition' in the toric case.

In this paper we are interested in hyperk\"ahler manifolds that are cones over 3-Sasakian manifolds.
For hypertoric cases this can be stated in terms of the hyperplanes as follows \cite{Bielawski2000}:
$M(\vec{u},\vec{0})$ is the metric cone over a compact 3-Sasakian manifold $X$ iff
(i) every subset of $\vec{u}$ with $d$ elements is linearly independent and 
(ii) every subset of $\vec{u}$ with less than $d$ elements is part of a $\Z$-basis of $\Z^d$. 

The proof of the above statement goes along the following lines \cite{Bielawski2000}:
Recall that $\H^n$ is the metric cone over the standard sphere $S^{4n-1}$, which is a 3-Sasakian manifold. Setting $\vec{\lambda} = \vec{0}$ allows one to induce a 3-Sasakian structure on $X$ from that on $S^{4n-1}$. Setting $\vec{\lambda} = \vec{0}$ means that all the hyperplanes intersect in the origin. This clearly violates the smoothness condition discussed above, but this is expected since we have a singularity at the tip of the cone. The other conditions on the hyperplanes guarantee that this is the only singularity.

Note that we are interested in 8-dimensional cones with hypertoric $T^2$ action, i.e. the case when $d=2$. In this case the condition (i) can be restated as the $\{u_k\}$ being pairwise linearly independent. Condition (ii) follows from the $u_k$ being primitive.

Thus the hypertoric $C(X)$ we are interested in can be described by a set of hyperplanes $H_k$ in $\R^2$ (i.e. lines) with normal vectors $u_k$, all passing through the origin\footnote{Technically the $H_k$ live in $\R^2 \times \R^2 \times\R^2$ but since all the $\lambda$'s are the same we can just look at a single $\R^2$. This is a simple case of the more general `canonical slice' discussed in \cite{Bullimore2016}.}.

We will now turn to our example of $C(\S)$.

\subsection{The cone $C(\S)$ as a hypertoric variety}

In this subsection we will realise the cone of our manifold, i.e. the Swann bundle $C(\S)=\U(Gr_2(\C^3))$, as a hypertoric variety and describe it in terms of a hyperplane arrangement.

In section \ref{sec:TheManifoldS} we obtained the Swann bundle $C(\S)=\U(Gr_2(\C^3))$ as a hyperk\"ahler quotient of $\H^3$ by $U(1)_Q$ with charges $[1,1,1;-1,-1,-1]$. The residual $T^2$ action on the quotient is generated by the $U(1)$'s $e_1$ and $e_2$ with charges $[1,0,0;-1,0,0]$ and $[0,1,0;0,-1,0]$ respectively.

The exact sequence \eqref{eq:exactseq1} in this case is
\begin{equation}
\begin{tikzcd}[column sep=small]
0 \arrow{r} &  \Z \arrow{r}{\iota} & \Z^3 \arrow{r}{\beta} & \Z^2 \arrow{r} & 0 
\end{tikzcd}  \label{eq:exactseq3}
\end{equation}
where we may represent $\iota$ by the matrix 
\begin{equation}
Q = \begin{pmatrix} 1 \\ 1 \\ 1 \end{pmatrix}
\end{equation}
and $\beta$ by
\begin{equation}
\tilde{Q}=\begin{pmatrix} 1 & 0 & -1\\ 0 & 1 & -1 \end{pmatrix} \, . \label{eq:Qtilde}
\end{equation}
From this we can read off the normal vectors of the hyperplanes to be 
\begin{equation}
u_1 = \begin{pmatrix} 1 \\ 0 \end{pmatrix}, 
u_2 = \begin{pmatrix} 0 \\ 1 \end{pmatrix},
u_3 = \begin{pmatrix} -1 \\ -1 \end{pmatrix} \, .
\end{equation}

\begin{figure}[!htbp]
\centering
\begin{tikzpicture}[scale=.5]
\draw [->] (-5,0) -- (5,0) node [below] {\small$x$};
\draw [->] (0,-5) -- (0,5) node [left] {\small$y$};
\draw [line width=0.5ex] (-4,0) -- (4,0);
\draw [line width=0.5ex] (0,-4) -- (0,4);
\draw [line width=0.5ex] (-3,3) -- (3,-3);
\end{tikzpicture}
\caption{Hyperplane arrangement for $C(\S)=\U(Gr_2(\C^3))$. The axes correspond to the components of the real moment map of the residual $T^2$-action.}
\label{fig:hyperplanesCS}
\end{figure}
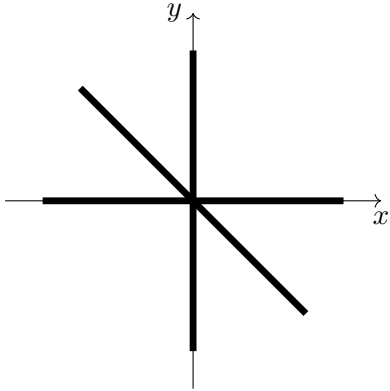

We get that all $a_k$'s of \eqref{eq:mRsubTorus} and \eqref{eq:mCsubTorus} are equal to one and since we take the zero level-sets of our moment maps all $c_i$'s are zero. Thus $\vec{\lambda}=\vec{0}$ and we can draw the hyperplane arrangement in $\R^2$ as in Figure \ref{fig:hyperplanesCS}.

Since $\vec{\lambda}=0$ and the $u_i$ are pairwise linearly independent, we verify that our manifold $C(\S)$ is indeed the cone over a compact 3-Sasakian manifold.

\section{Calculating the superdeterminant} \label{sec:HoloFunCalc} 

In this section we will compute the superdeterminant \eqref{eq:sdet}, and hence the perturbative partition function of our SYM theory on $\S$, by considering holomorphic functions on the cone $C(\S)$. We will then verify the result by an index calculation. 
This is similar in spirit to how the 5D index calculations of Qiu and Zabzine in \cite{Qiu:2013pta} were interpreted in terms of toric data by Schmude in \cite{Schmude:2014lfa}.

\subsection{Holomorphic functions on $C(\S)$}
We now follow \cite[Section~6.1]{Bullimore2016}\footnote{We thank Nicholas Proudfoot for pointing out this reference to us.} to obtain a description of the holomorphic functions of $C(\S)$ from its hyperplane arrangement. 

Since $C(\S)$ is hyperk\"ahler the term `holomorphic' is ambiguous unless we specify a complex structure. Here we will use the complex structure $I$ in \eqref{eq:acsI}.

The generators of the ring of holomorphic functions come in two types, those that are neutral under the $T^2$ action and those that are charged. The neutral generators in our example are given by $N_1=z_1w_1$ and $N_2=z_2w_2$. To describe the charged ones we use the normal vectors of the hyperplanes, arranged into the matrix $\tilde{Q}$ in \eqref{eq:Qtilde}. For every element $A\in \Z^2$ of the $T^2$ charge lattice, define
\begin{align}
C^A := \prod_{i=1}^3 
\begin{cases}
z_i^{|\tilde{Q}^i_A|}, & \text{if}\ \tilde{Q}^i_A>0 \\
w_i^{|\tilde{Q}^i_A|}, & \text{if}\ \tilde{Q}^i_A<0 
\end{cases} \, ,
\label{eq:ringrelations}
\end{align}
where $\tilde{Q}_A := \tilde{Q}^T A \in \Z^3$.
In our case $\tilde{Q}_A = (A_1,A_2,-A_1-A_2)$.
The $C^A$ obey the ring relations \cite{Bullimore2016}
\begin{align}
C^A C^B = C^{A+B} \prod\limits_{i \text{ s.t. } \tilde{Q}_A^i\tilde{Q}_B^i < 0 } (z_iw_i)^{\text{min}(|\tilde{Q}^i_A|,|\tilde{Q}^i_B|)} \, .
\end{align}
These $C^A$ and the $N_i= z_iw_i$ together generate the ring of holomorphic functions \cite{Bullimore2016}.

\begin{figure}[!htbp]
\centering
\begin{tikzpicture}[scale=.7]
\draw [->] (-5,0) -- (5,0) node [below] {\small$x$};
\draw [->] (0,-5) -- (0,5) node [left] {\small$y$};
\draw [line width=0.25ex] (-4,0) -- (4,0);
\draw [line width=0.25ex] (0,-4) -- (0,4);
\draw [line width=0.25ex] (-4,4) -- (4,-4);
\foreach \x in {-3,...,3}
\foreach \y in {-3,...,3}
	{
	\draw [black,fill=black] (\x,\y) circle (0.5ex);
	 }
\node at (4,4) {\large I};
\node at (4,-2) {\large II};
\node at (2,-4) {\large III};
\node at (-4,-4) {\large IV};
\node at (-4,2) {\large V};
\node at (-2,4) {\large VI};
\end{tikzpicture}
\caption{Hyperplanes superimposed on $T^2$ charge lattice.}
\label{fig:hyperplanesChargeLattice}
\end{figure}
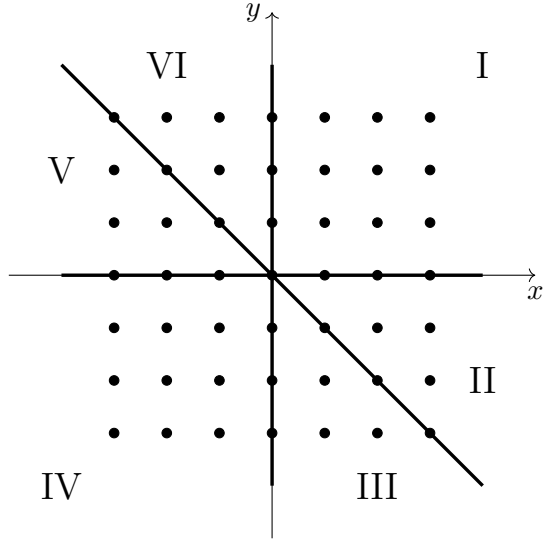

This can be interpreted geometrically by drawing the hyperplanes in the $T^2$ charge lattice as in Figure \ref{fig:hyperplanesChargeLattice}.
We see that the hyperplanes cut the lattice into six regions which we have labelled by roman numerals in Figure \ref{fig:hyperplanesChargeLattice}.
For two points $A,B$ in this lattice, the ring relations say that $C^A C^B$ is $C^{A+B}$ up to a correction factor for each hyperplane between the rays $\rho(A)$ and $\rho(B)$. In particular, if they lie in the same region then there are no corrections. A finite set of generators for the ring of holomorphic functions is then given by the $N_i$ together with $\{ C^A\}_{A \in \mathcal{A}}$ such that the $A \in \mathcal{A}$ generate the integral lattice inside each of the regions \cite{Bullimore2016}.

In our example, the generators are thus $N_1=z_1w_1$ and $N_2=z_2w_2$ together with
\begin{align}
C^{(1,0)}=z_1w_3, & & C^{(0,1)}=z_2w_3, & & C^{(1,-1)}=z_1w_2, \\ \notag
C^{(-1,0)}=w_1z_3, & & C^{(0,-1)}=w_2z_3, & & C^{(-1,1)}=w_1z_2.
\end{align}
We can now enumerate the holomorphic functions as follows. For each point $A$ in the charge lattice we get $C^A$ along with $N_1^m N_2^n$ for any $m,n=0,1,2,\dots$.

In Region I we write each lattice point $A = a (0,1) + b(1,0)$ and get
\begin{align}
(C^{(0,1)})^a (C^{(1,0)})^b N_1^m N_2^n = (z_2w_3)^a (z_1w_3)^b (z_1w_1)^m (z_2w_2)^n = z_1^{b+m} w_1^m z_2^{a+n}w_2^{n}z_3^{0}w_3^{a+b} \, .
\end{align}

In Region II we write each lattice point $A = a (1,0) + b(1,-1)$ and get
\begin{align}
(C^{(1,0)})^a (C^{(1,-1)})^b N_1^m N_2^n = (z_1w_3)^a (z_1w_2)^b (z_1w_1)^m (z_2w_2)^n = z_1^{a+b+m} w_1^m z_2^{n}w_2^{b+n}z_3^{0}w_3^{a}  \, .
\end{align}
We take $a=0,1,2, \dots$ and $b=1,2,3, \dots$ to avoid double-counting the borders between the regions.

Continuing like this we get the functions listed in Table \ref{tab:holofun}.

\subsection{Calculation of superdeterminant}

From the holomorphic functions on the cone $C(X)$ we can now calculate the superdeterminant \eqref{eq:sdet}. Having picked the complex structure $I$ we will use the associated Reeb vector $R_1$ of \eqref{eq:Reeb1}. The corresponding $U(1)$ action of this Reeb has charge $[1,1,1;1,1,1]$ and we will denote its equivariant parameter by $\mu$. 
There are three other $U(1)$'s, $e_{1,2,3}$, with charges 
\begin{align}
e_1: [1,0,0;-1,0,0], e_2 : [0,1,0;0,-1,0], e_3 : [0,0,1;0,0,-1] \, , \label{eq:e123charges}
\end{align}
whose equivariant parameters we will denote by $\w_{1,2,3}$. Note that $e_1+e_2+e_3 \sim 0$ since this combination is proportional to $Q=[1,1,1;-1,-1,-1]$.

In the previous subsections we found the holomorphic functions in terms of hyperplanes that formed six regions in Figure \ref{fig:hyperplanesChargeLattice}. These are given in Table \ref{tab:holofun} along with their $U(1)$ weights.
\begin{table}[!htbp]
\begin{tabular}{|c|cccccc|c|c|c|}
\hline
{Region}  &  \multicolumn{6}{c|}{Functions}  & {$\w_1$} & {$\w_2$} & {$\w_3$} \\
\hline
{I}  & $ z_1^{b+m} $ & $  w_1^m $ & $ z_2^{a+n} $ & $ w_2^{n} $ & $ z_3^{0} $ & $ w_3^{a+b}$
& $b$ & $a$ & $-a-b$ \\
{II}  & $ z_1^{a+b+m} $ & $  w_1^m $ & $ z_2^{n} $ & $ w_2^{b+n} $ & $ z_3^{0} $ & $ w_3^{a}$ & $a+b$ & $-b$ & $-a$ \\
{III} & $ z_1^{a+m} $ & $  w_1^{m} $ & $ z_2^{n} $ & $ w_2^{a+b+n} $ & $ z_3^{b} $ & $ w_3^{0}$ & $a$ & $-a-b$ & $b$ \\
{IV}  & $ z_1^{m} $ & $ w_1^{b+m} $ & $ z_2^{n} $ & $ w_2^{a+n} $ & $ z_3^{a+b} $ & $ w_3^{0}$ & $-b$ & $-a$ & $a+b$ \\
{V}  & $ z_1^{m} $ & $ w_1^{a+b+m} $ & $ z_2^{b+n} $ & $ w_2^{n} $ & $ z_3^{a} $ & $ w_3^{0} $ & $-a-b$ & $b$ & $a$ \\
{VI}  & $ z_1^{m} $ & $ w_1^{a+m} $ & $ z_2^{a+b+n} $ & $ w_2^{n} $ & $ z_3^{0} $ & $ w_3^{b}$ & $-a$ & $a+b$ & $-b$ \\
{O}  & $ z_1^{m} $ & $ w_1^{m} $ & $ z_2^{n} $ & $ w_2^{n} $ & $ z_3^{0} $ & $ w_3^{0}$ & $0$ & $0$ & $0$  \\
\hline
\end{tabular}
\caption{Holomorphic functions on $C(\S)$ and their weights under $\w_{1,2,3}$. All functions have $\mu$-weight $2(a+b+m+n)$.  Here $m,n,a=0,1,2, \dots$, but $b=1,2,3, \dots$, to avoid overcounting the borders between the regions. We have included the origin $O$ separately (where $a=b=0$).}
\label{tab:holofun}
\end{table}
From Table \ref{tab:holofun} we read off the contribution to the superdeterminant coming from $H_{KR}^{(0,0)}$:
\begin{align}
\prod_{m,n,a=0, b=1}^\infty \Big( & \left( x+ (b)\w_1 + (a)\w_2 +(-a-b)\w_3 +2(a+b+m+n)\mu   \right) \notag\\
&\left( x+ (a+b)\w_1 + (-b)\w_2 +(-a)\w_3 +2(a+b+m+n)\mu  \right)\notag\\
&\left( x+ (a)\w_1 + (-a-b)\w_2 + (b)\w_3 +2(a+b+m+n)\mu   \right) \notag\\
&\left( x+ (-b)\w_1 + (-a)\w_2 + (a+b)\w_3 +2(a+b+m+n) \mu   \right) \label{eq:hypertoricAns} \\
&\left( x+ (-a-b)\w_1 + (b)\w_2 + (a)\w_3 +2(a+b+m+n)\mu   \right)  \notag\\
&\left( x+ (-a)\w_1 + (a+b)\w_2 + (-b)\w_3 +2(a+b+m+n)\mu   \right)  \Big) \notag\\
\times \prod_{m,n=0}^\infty & \left( x +2(m+n) \mu  \right) \, . \notag
\end{align}
Techincally the factor of $x$ in this expression, corresponding to the constant monomial, should not be included here. It arises when one uses the Weyl integration formula later on. But for ease of notation we include it already at this stage.

\begin{figure}[htbp!]
\centering
\tdplotsetmaincoords{60}{110}
\begin{tikzpicture}[scale=3,tdplot_main_coords]
    \draw[thick,->] (0,0,0) -- (1,0,0) node[anchor=north east]{$i$};
    \draw[thick,->] (0,0,0) -- (0,1,0) node[anchor=north west]{$j$};
    \draw[thick,->] (0,0,0) -- (0,0,1) node[anchor=south]{$k$};
	
	\coordinate (O) (0,0,0);
	\coordinate (IJ0) (1,1,0);
	\coordinate (IJ1) (1,1,1);
    \filldraw[draw=red,fill=red!50,fill opacity=0.5]
              (0,0,0)
            -- (1.0,1.0,0) node[right,color=red,text opacity=1.0]{$i=j$}
            -- (1.0,1.0,1)
            -- (0,0,1)
            -- cycle;
    \filldraw[draw=green,fill=green!50,fill opacity=0.5, text opacity=1.0]
              (0,0,0)
            -- (1.0,0,1.0) node[left,color=green]{$i=k$}
            -- (1.0,1.0,1.0) 
            -- (0,1.0,0)
            -- cycle;
    \filldraw[draw=blue,fill=blue!50,fill opacity=0.5,text opacity=1.0]
              (0,0,0)
            -- (0,1.0,1.0) node[right,color=blue]{$j=k$}
            -- (1,1.0,1.0) 
            -- (1,0,0)
            -- cycle ;
    \draw[thin,->,gray] (0,0,0) -- ({1.2},{1.2},{1.2}) node[right,color=gray,text opacity=1.0] {$i=j=k$};
\end{tikzpicture}
\caption{Sketch of the positive octant of $\mathbb{Z}^3$ cut into six pieces by the planes $\{i=j\}, \{i=k\}, \{j=k\}$.}
\label{fig:planesCuttingZ3}
\end{figure}
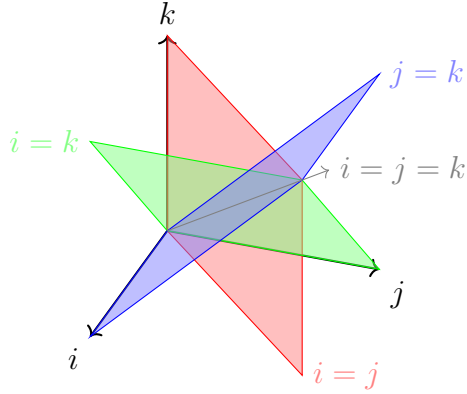

We can rewrite \eqref{eq:hypertoricAns} in terms of the variables $\w_{ab} = \w_a-\w_b$ as follows:
\begin{align}
\prod_{n=1}^\infty \Bigg( 
&\left(\prod_{i<k \leq j \leq n}(x+i\w_{31}+j\w_{23}+k\w_{12}+2n\mu)\right) \notag\\
&\left(\prod_{i \leq j <k \leq n}(x+i\w_{31}+j\w_{23}+k\w_{12}+2n\mu)\right) \notag\\
&\left(\prod_{j<i \leq k \leq n}(x+i\w_{31}+j\w_{23}+k\w_{12}+2n\mu)\right) \notag\\
&\left(\prod_{j \leq k < i \leq n}(x+i\w_{31}+j\w_{23}+k\w_{12}+2n\mu)\right) \\
&\left(\prod_{k<j \leq i \leq n}(x+i\w_{31}+j\w_{23}+k\w_{12}+2n\mu)\right) \notag\\
&\left(\prod_{k \leq i < j \leq n}(x+i\w_{31}+j\w_{23}+k\w_{12}+2n\mu)\right) \Bigg) \notag \\
\times \prod_{n=0}^\infty & \left( \prod_{i=j=k \leq n} (x+ i\w_{31} + j\w_{23} + k\w_{12} + 2n\mu)  \right) \, . \notag
\end{align}
This expression simplifies to 
\begin{align}
\prod_{n=0}^{\infty} \prod_{i,j,k=0}^n (x + i\w_{31}+j\w_{23}+k\w_{12} + 2n\mu) \, . \label{eq:hypertoricAnsRewrite}
\end{align}
This can be seen by realising that the first six products correspond to the regions obtained by cutting the positive octant of $\mathbb{Z}^3$ into six pieces by the planes $\{i=j\}, \{i=k\}, \{j=k\}$, and the seventh product corresponds to their intersection in the line $\{i=j=k\}$. This is illustrated in Figure \ref{fig:planesCuttingZ3}.

Combining \eqref{eq:hypertoricAnsRewrite} with the shifted contribution from $H_{KR}^{(0,3)}$ we get
\begin{align}
sdet_{\Omega_H^{(0,\bullet)}}(-\L_R + x) \sim 
\frac{ \prod\limits_{n=0}^\infty
\prod\limits_{i,j,k=0}^n (i\w_{31}+j\w_{23}+k\w_{12}+2n\mu +x )}{\prod\limits_{n=0}^\infty
\prod\limits_{i,j,k=0}^n (-i\w_{31}-j\w_{23}-k\w_{12}-2(n+2)\mu +x)} \, .
\end{align}
Noting that the products above are invariant under $\w_{ij} \rightarrow -\w_{ij}$ we rewrite it as
\begin{align}
sdet_{\Omega_H^{(0,\bullet)}}(-\L_R + x) \sim 
\frac{ \prod\limits_{n=0}^\infty
\prod\limits_{i,j,k=0}^n (i\w_{31}+j\w_{23}+k\w_{12}+2n\mu +x)}{\prod\limits_{n=2}^\infty
\prod\limits_{i,j,k=0}^{n-2} (i\w_{31}+j\w_{23}+k\w_{12}+2n \mu-x)} \,. \label{eq:holofunAnswer}
\end{align}

The perturbative partition function for maximally supersymmetric Yang-Mills on our 7D manifold $\S$ is thus given by
\begin{align}
Z^{\text{pert}}_{\S}
&= \int\limits_t d \sigma \,  e^{-\frac{24}{g^2_7} V_7 \Tr(\sigma^2)} \prod_\beta S( i \bra \sigma, \beta \ket |\mu,\w_1,\w_2,\w_3)\, , \label{pertparfunS}
\end{align}
where $\beta$ are the roots of the Lie algebra $g$ with Cartan subalgebra $t$ and the special function $S$ is given by the infinite product formula\footnote{As they stand, these infinite products are divergent, but we can make them finite by $\zeta$-function regularisation, see appendix \ref{AppAsymp}.} 
\begin{align}
S(x|\mu,\w_1,\w_2,\w_3) &=
\frac{ \prod\limits_{n=0}^\infty
\prod\limits_{i,j,k=0}^n (i\w_{31}+j\w_{23}+k\w_{12}+2n\mu +x)}{\prod\limits_{n=2}^\infty
\prod\limits_{i,j,k=0}^{n-2} (i\w_{31}+j\w_{23}+k\w_{12}+2n \mu-x)} \,. \label{eq:specFunDefnS}
\end{align}

For the `unsquashed' geometry where all $\w_i,\mu=1$ we get
\begin{align}
S(x) &=\frac{\prod \limits_{n=0}^\infty (2n+x)^{(n+1)^3}}{\prod\limits_{n=2}^\infty (2n-x)^{(n-1)^3}} =x (2+x)^{8} \prod \limits_{n=2}^\infty \frac{(2n+x)^{(n+1)^3}}{ (2n-x)^{(n-1)^3}} \,. \label{eq:specFunSUnsquashed}
\end{align}
In the matrix model for the `unsquashed' case we can further use the reflection symmetry, $\beta \rightarrow -\beta$, in the product over the roots to get
\begin{align}
\prod_\beta S( i \bra \sigma, \beta \ket ) = \prod_\beta ( i \bra \sigma, \beta \ket) \prod \limits_{n=1}^\infty (2n + i \bra \sigma, \beta \ket)^{6n^2+2} \,.
\end{align}
These formulas are in analogy with those discussed for the `unsquashed' spheres in \cite{Minahan:2015jta}.

In appendix \ref{sec:IndexCalc} we verify the result \eqref{eq:holofunAnswer} by an index calculation. More specifically, we compute the superdeterminant as the index of the $\bar{\partial}_H$-operator w.r.t. the Reeb $R_1$ in \eqref{eq:Reeb1}. The approach follows closely that in \cite{Qiu:2013pta} and serves as a double check of the hypertoric method presented above.

\section{Factorisation} \label{sec:factor}

In the previous section we found the \emph{perturbative} partition function for 7D SYM on the 3-Sasakian manifold $\S$. It was given in terms of a matrix model involving a special function. Finding the \emph{full} partition function is beyond us at the moment, but by studying factorisation properties of the perturbative answer one can make some guesses. For example, the perturbative answer for $S^7$ is given in terms of a quadruple sine function that can be factorised into four pieces:
\begin{align} \label{eq:factorS7Maintext}
S_4 (x|\w_1,\w_2\,\w_3,\w_4) =  e^{\frac{\pi i }{4!} B_{4,4}(x|\w_1,\w_2\,w_3,\w_4)}\prod_{k =1}^4 (z_k|\qvec_k)_\infty \, .
\end{align}
Here $B_{4,4}$ is a multiple Bernoulli polynomial and $(z_k|\qvec_k)_\infty$ are $q$-shifted factorials, see appendix \ref{AppAsymp} for details. The important point is that each piece on the RHS can be interpreted as the perturbative Nekrasov partition function on $S^1 \times_\epsilon \C^3$ \cite{Nekrasov2008}. It is then natural to conjecture that the full partition function for  $S^7$ enjoys the same factorisation in terms of full Nekrasov partition functions, as was argued in \cite{Minahan:2015jta}.

Similar arguments were used in \cite{Polydorou2017} for 7D toric Sasaki-Einstein manifolds. These were based on analogous factorisation results for generalised quadruple sine functions  \cite{Winding:2016wpw} and could be interpreted in terms of toric data.

In \cite{Polydorou2017} another factorisation of the perturbative answer for $S^7$ was discussed, based on the 3-Sasakian structure of $S^7$. It was argued that the quadruple sine function \eqref{eq:S4Defn} could be written as an infinite product over two double sine functions. This corresponds to viewing $S^7$ as an $SU(2)$-bundle over $S^4$ via the quaternionic Hopf fibration. There are two fixed points, each giving a double sine corresponding to the $SU(2)$ fibre.

The 3-Sasakian manifold $\S$ studied in this paper can be seen as an $SO(3)$-bundle over $\CP^2$. It is therefore natural to ask if we can make similar statements here, i.e. find a factorisation of the special function \eqref{eq:specFunDefnS} in terms of $SO(3)$ data. The approach we take  here is based on the index calculation and details can be found in appendix \ref{app:sdetSO3}. It would be interesting if the same result could also be seen from the hypertoric viewpoint of section \ref{sec:HoloFunCalc}.

In viewing our manifold $\S$ as an $SO(3)$-fibration over $\CP^2$ there are three fixed points contributing to the index (see appendix \ref{sec:IndexCalc}). At each fixed point we get a contribution from the fibre and the normal bundle. The contribution from the fibre can be written in terms of $SO(3)$-characters whose infinite product expression correspond to an `odd double sine function'. Recall that the regular double sine function is given by the infinite product 
\begin{equation}
S_2(x|\w_1,\w_2) = \frac{\prod\limits_{\substack{k,l=0}}^{\infty} \left( x + k \w_1 + l\w_2 \right) }{\prod\limits_{\substack{k,l=1}}^{\infty} \left( -x + k \w_1 + l\w_2  \right)} \, .
\end{equation}
If we restrict the products to the sublattice of $\mathbb{Z}_{\geq 0}^2$ where $k+l$ is odd we get what we refer to as the `odd double sine', $S_2^{odd}$.

Using this we obtain the following expression for the superdeterminant (see appendix \ref{app:sdetSO3} for details):
\begin{align}
  sdet_{\Omega_H^{(0,\bullet)}}(-\L_R + x) \sim &  \left( \prod_{i=1, j=0} S_2^{odd} \left( x+ i\w_{23}+ j\w_{13} - \mu +\tfrac12 \w_{12} | \mu -\tfrac12 \w_{12}, \mu + \tfrac12 \w_{12}\right)^{-1} \right)  \notag \\ 
&  \left(  \prod_{i=0, j=0} S_2^{odd} \left( x+ i\w_{12}+ j\w_{23} - \mu +\tfrac12 \w_{13} | \mu -\tfrac12 \w_{13}, \mu + \tfrac12 \w_{13}\right) \right) \label{eq:factorisationI}\\ 
&  \left( \prod_{i=1, j=0} S_2^{odd} \left( x+ i\w_{12}+ j\w_{13} - \mu +\tfrac12 \w_{23} | \mu -\tfrac12 \w_{23}, \mu + \tfrac12 \w_{23}\right)^{-1} \right)  \,. \notag
\end{align}
Having calculated the same superdeterminant in two ways, we obtain the factorisation of the special function in \eqref{eq:specFunDefnS}:
\begin{align}
& \frac{ \prod\limits_{n=0}^\infty
\prod\limits_{i,j,k=0}^n (i\w_{31}+j\w_{23}+k\w_{12}+2n\mu +x)}{\prod\limits_{n=2}^\infty
\prod\limits_{i,j,k=0}^{n-2} (i\w_{31}+j\w_{23}+k\w_{12}+2n \mu-x)}  \notag  \\
&\quad =  \left( \prod_{i=1, j=0} S_2^{odd} \left( x+ i\w_{23}+ j\w_{13} - \mu +\tfrac12 \w_{12} | \mu -\tfrac12 \w_{12}, \mu + \tfrac12 \w_{12}\right)^{-1} \right)  \notag \\ 
&\qquad  \left(  \prod_{i=0, j=0} S_2^{odd} \left( x+ i\w_{12}+ j\w_{23} - \mu +\tfrac12 \w_{13} | \mu -\tfrac12 \w_{13}, \mu + \tfrac12 \w_{13}\right) \right) \label{eq:factorisationFunI}\\ 
& \qquad \left( \prod_{i=1, j=0} S_2^{odd} \left( x+ i\w_{12}+ j\w_{13} - \mu +\tfrac12 \w_{23} | \mu -\tfrac12 \w_{23}, \mu + \tfrac12 \w_{23}\right)^{-1} \right)  \, , \notag
\end{align}
where we have ignored Bernoulli-type factors (see appendix \ref{AppAsymp} for details). Such factorisations are usually proved using integral representations of the functions involved, but since we lack such descriptions we instead take the more pedestrian route of comparing poles, zeros and asymptotics. That the poles and zeros of the expressions in \eqref{eq:factorisationFunI} match follows from that they describe the same superdeterminant, but it can also be verified by a direct computation. The asymptotic behaviour of these functions is a bit involved and we refer to appendix \ref{AppAsymp} for details.

In \cite{Polydorou2017} it was argued that the perturbative partition function for $S^7$ could be written as two copies of the perturbative partition function on $S^3 \times \C^2$. This corresponded to viewing $S^7$ as an $SU(2)$-bundle over $S^4$, with each of the two fixed points giving a double sine $S_2$.
The corresponding statment here would be that the perturbative partition function for $\S$ can be written as three copies of the perturbative partition function on $\mathbb{R}P^3 \times \C^2$. This is based on viewing $\S$ as an $SO(3)$ fibration over $\CP^2$, with each of the three fixed points giving an odd double sine $S_2^{odd}$.
However, these statements are speculatory and one would need to first understand what the Nekrasov partition functions for $S^3 \times \C^2$ and $\mathbb{R}P^3 \times \C^2$ are in order to make  more precise statements.

In figure \ref{fig:factorisation} we give an illustration of the factorisation results for $S^7$ and $\S$.  In the spirit of toric diagrams these pictures attempt to illustrate the collapsing of tori along the edges of the polygons. The corners of the polygons correspond to fixed points where all tori collapse.

\begin{figure}[htbp!]
\centering
\caption{Cartoons illustrating the different factorisation results for $S^7$ and $\S$.}
\begin{subfigure}[t]{0.45\textwidth}
\caption{$S^7:$}
\begin{tikzpicture}
\def\r{0.25}
\draw [black, xshift=0cm, name path=one] plot [smooth, tension=1] coordinates { (-1,0) (0,0.5) (1,0) };
\draw [black, xshift=0cm, name path=two] plot [smooth, tension=1] coordinates { (-1,0) (0,-0.5) (1,0) };
\tikzfillbetween[of=one and two,split] {gray, opacity=0.5};
\draw (-1-\r,0) circle (\r);
\draw (1+\r,0) circle (\r);
\node at (-2,0) {$S^3$};
\node at (2,0) {$S^3$};
\end{tikzpicture}
\end{subfigure}
\begin{subfigure}[t]{0.45\textwidth}
\caption{$\S:$}
\begin{tikzpicture}
\def\r{0.25}
\draw [black, xshift=0cm, name path=one] plot [smooth, tension=1] coordinates { (-1,0) (1,0) };
\draw [black, xshift=0cm, name path=two] plot [smooth, tension=1] coordinates { (-1,0) (0,{sqrt(3)}) };
\draw [black, xshift=0cm, name path=two] plot [smooth, tension=1] coordinates { (1,0) (0,{sqrt(3)}) };
\tikzfillbetween[of=one and two,split] {gray, opacity=0.5};
\draw (-1-0.2165,0-0.125) circle (\r);
\draw (1+0.2165,0-0.125) circle (\r);
\draw (0,1.7320+\r) circle (\r);
\node at (-2,0) {$\mathbb{R}P^3$};
\node at (0,2+2*\r) {$\mathbb{R}P^3$};
\node at (2,0) {$\mathbb{R}P^3$};
\end{tikzpicture}
\end{subfigure}
\label{fig:factorisation}
\end{figure}

\section{Summary}\label{sec:summary} 

In this paper we have considered 7D maximally supersymmetric Yang-Mills on a specific 7D 3-Sasakian manifold $\S$. Since the manifold is Sasaki-Einstein the localisation procedure of \cite{Polydorou2017} could be applied and the perturbative partition function written in terms of a superdeterminant. This superdeterminant can be calculated in terms of the holomorphic functions on the metric cone of the manifold. For \emph{toric} Sasaki-Einstein manifolds (hamiltonian $T^4$ action on the cone) this count has been obtained in terms of toric data in \cite{Polydorou2017} in an analogous way to the 5D case. In this paper we have considered a manifold that is \emph{not} toric but whose metric cone has \emph{hypertoric} symmetry (hyperhamiltonian $T^2$ action on the cone). The hypertoric structure allowed us to enumerate the holomorphic functions and thereby compute the perturbative partition function. The answer was given in terms of a matrix model involvning a special function. This special function was then factorised into odd double sine functions which  corresponded to viewing the manifold as an $SO(3)$-bundle over $\CP^2$.

The explicit example in this paper opens up for further studies of the role of 3-Sasakian and hypertoric structures in 7D SYM. It should be relatively straight-forward to generalise the arguments presented here to any given 3-Sasakian manifold whose metric cone is hypertoric. The challenge would be to obtain a closed form answer in terms of hypertoric data that would work for aribtrary such manifolds. In other words, it would be interesting to obtain the answer as a special function defined in terms of hypertoric data, analogous to the generalised multiple sine functions in the toric case.

In general, a 3-Sasakian manifold has a three-dimensional foliation whose generic leaves are either $SU(2)$ or $SO(3)$, see \cite{Boyer1998}. It is natural to guess that this leads to  factorisations such as those discussed in section \ref{sec:factor}. It would be interesting to explore this further and also see what role, if any, the hypertoric structure plays for this. Maybe it could also provide some hints about the Nekrasov partition functions for $S^3 \times \C^2$ and $\mathbb{R}P^3 \times \C^2$.

\vspace{1cm}

{\bf Acknowledgements}\\
The author is very grateful to Jian Qiu and Maxim Zabzine for many useful discussions, suggestions and general supervision of this work. This research is supported in part by the grant ``Geometry and Physics'' from the Knut and Alice Wallenberg foundation.


\appendix 

\section{Hyperk\"ahler structure of flat quaternionic space} \label{AppQuaternions}

In this appendix we give som details about our conventions, notation, and basic properties of the flat quaternionic space $\H$.

Consider the quaternionic vector space $\H$ with coordinates $q = q_0 + i q_1 + j q_2 + k q_3$. Here $q_{0,1,2,3} \in \R$ and $i,j,k$ are the unit quaternions with
\begin{equation}
i^2=j^2=k^2=ijk =-1\, .
\end{equation}
 We will often map $\H$ to $\C \times (\C)^*$ via  $q=z+ j w$, where $z=q_0+iq_1$ and $w=q_2-iq_3$.

%

We will take the three complex structures $I_a$ on $\H$ to be given by left multiplication by $e_a= i, j, k$. 

The three symplectic forms are parametrised as
\begin{equation}
\w_a = \frac{1}{2} dq e_a d\qb.
\end{equation}
More explicitly
\begin{align}
\w_1 &= \frac{1}{2} dq i d\qb = \frac{1}{2} (dq_0 + idq_1 + jdq_2 + kdq_3) i (dq_0 - idq_1 -jdq_2 - k dq_3) \\
&= dq_0 dq_1 - dq_2 dq_3 = \frac{i}{2}(dz d\zb +dw d\wb )\, ,
\end{align}
and similarly
\begin{align}
\w_2 &= dq_0 dq_2 + dq_1 dq_3 = \frac{1}{2}(dz dw + d\zb d\wb)  \, ,\\
\w_3 &= dq_0 dq_3 - dq_1 dq_2 = \frac{i}{2}(dz dw -d\zb d\wb).
\end{align}
The vector fields corresponding to left-multiplication by $e_a= i,j,k$ are given by $V_a = (e_a q)^i \partial_{q^i}$, that is
\begin{align}
V_1 &= -q_1 \partial_0 + q_0 \partial_1 -q_3 \partial_2 +q_2 \partial_3 = iz\partial_z - iw \partial_w + c.c. \, , \\
V_2 &= -q_2 \partial_0 + q_3 \partial_1 +q_0 \partial_2 -q_1 \partial_3 = -w\partial_z + z \partial_w + c.c. \, , \\
V_3 &= -q_3 \partial_0 - q_2 \partial_1 +q_1 \partial_2 +q_0 \partial_3 = -i w \partial_z -iz \partial_w  + c.c. \, .
\end{align}
Their moment maps are given by
\begin{equation}
\mu_a = -\frac12 \qb e_a q.
\end{equation}

Let us focus on the action of left-multiplication by $i$. The moment map is then
\begin{align}
\mu_1 &= -\frac12 ( q_0 - i q_1 - j q_2 - k q_3) i ( q_0 + i q_1 + j q_2 + k q_3) \\
&= \frac{i}{2}(-q_0^2-q_1^2+q_2^2+q_3^2) + j(q_0q_3-q_1q_2) + k(-q_0q_2-q_1q_3)\\
&= i\mu_{11} + j\mu_{12} + k\mu_{13} \, ,
\end{align}
and a direct computation shows that
\begin{align}
\iota_{V_1} \w_1 &= d (\tfrac12 (-q_0^2-q_1^2+q_2^2+q_3^2)) = d\mu_{11}\\
\iota_{V_1} \w_2 &= d (q_0q_3-q_1q_2) = d \mu_{12}\\
\iota_{V_1} \w_3 &= d (-q_0q_2-q_1q_3) = d\mu_{13} \,.
\end{align}

In terms of the complex coordinates $z,w$ we write this as the real and complex moment maps
\begin{align}
\mu_1^{\R}  &:= \mu_{11} = \frac{1}{2}(-|z|^2+|w|^2) \, , \\
\mu_1^{\C} &:= \mu_{12}-i \mu_{13} = i z w \, .
\end{align}

\section{The index calculation} \label{sec:IndexCalc} 

In this section we will find the superdeterminant \eqref{eq:sdet} on the 3-Sasakian manifold $\S$ by computing the index of the $\db_H$-operator w.r.t. the Reeb $R_1$ in \eqref{eq:Reeb1}.
This is a transversally elliptic operator and a proper mathematical treatment of this topic can be found in the book by Atiyah \cite{AtiyahElliptic}. The parts needed for our calculation are reviewed in \cite{Qiu:2013pta} and we will follow their method for calculating the index. 
The general idea is to deform the symbol of the operator using vector fields so that the symbol becomes trivial except at the zeros of the vector field. At these fixed points the contribution to the symbol is computed `by hand' using the knowledge of the local geometry close to the fixed point. A nice and pedagogical review is found in appendix D of \cite{Qiu:2013pta}.

We will parametrise the manifold $\S$ as in section \ref{sec:TheManifoldS} and denote the $U(1)$'s available in the geometry by their charges
\begin{align}
e_1=[1,0,0;-1,0,0], e_2 = [0,1,0;0,-1,0], e_3 = [0,0,1;0,0,-1],  R_1=[1,1,1;1,1,1] \, , \label{eq:U1sofS}
\end{align}
and the corresponding angle variables by $\alpha, \beta, \gamma, \theta$. Note that $e_1 + e_2 + e_3 \sim 0$ since this combination is proportional to $Q=[1,1,1;-1,-1,-1]$. Here $\theta$ is the angle of the `Hopf fibre' but, as seen from \eqref{eq:SO3Fibre}, when $\theta=\pi$ its effect can be undone by $Q$ and hence the fibre is $SO(3)$. We will however work in terms of $SU(2)$ and then use the isomorphism $SU(2)/ \mathbb{Z}_2 \iso SO(3)$ for the final results.

We will consider the index of $\db_H$ w.r.t. the Reeb $R_1$ and use $e_1$ and $e_2$ to deform the symbol of this operator. The fixed points are the loci where $\epsilon_1 e_1 + \epsilon_2 e_2 \in \mathrm{span} \langle R_1, Q \rangle$. This happens when only one of the $z$'s and one of the $w$'s is non-zero, for example $\vz = (1,0,0), \vw = (0,1,0)$. There are six choices for this, but some of them lie in the same orbit under $SU(2)$, so we organise them into the three choices
\begin{align}
\begin{pmatrix} \vz \\ \vw \end{pmatrix} &= \begin{pmatrix} 1 & 0 & 0 \\ 0 & 1 & 0 \end{pmatrix}, \,
\begin{pmatrix} 0 & 0 & 1 \\ 1 & 0 & 0 \end{pmatrix}, \,
\begin{pmatrix} 0 & 1 & 0 \\ 0 & 0 & 1 \end{pmatrix} \, ,
\end{align}
corresponding to the points
\begin{align}
\vec{x} = [\vz \times \vw^*] = [0,0,1],\, [0,1,0],\,  [1,0,0] \, , \label{eq:fps}
\end{align}
of $\CP^2$.

We will first work in the neighbourhood of $\vec{x}=[0,0,1]$. We choose the coordinates of the holomorphic tangent space as
\begin{equation}
\delta \vz = a(0,0,1), \quad \delta \vw = b(0,0,1) \, ,
\end{equation}
and the transverse holomorphic coordinate of the $SU(2)$ fibre as
\begin{equation}
\begin{pmatrix} \vz \\ \vw \end{pmatrix} = \begin{pmatrix} 1 & -u & 0 \\ u & 1 & 0 \end{pmatrix} \frac{e^{i\theta}}{\epsilon_u}, \quad \epsilon_u^2 = 1+ |u|^2 \, . \label{eq:SU2Act1}
\end{equation}
If we use the quaternion embedding $q=z + jw \mapsto \begin{pmatrix}  z & -\bar{w} \\ w & \bar{z} \end{pmatrix}$ this action would correspond to right multiplication by the $SU(2)$ matrix
\begin{align}
\frac{1}{\epsilon_u}  \begin{pmatrix} e^{i\theta} & -\bar{u}e^{-i\theta} \\ ue^{i\theta} & e^{-i\theta} \end{pmatrix} \, .
\end{align}
For example, the second column in $\begin{pmatrix} \vz \\ \vw \end{pmatrix}  = \begin{pmatrix} 1 & 0 & 0 \\ 0 & 1 & 0 \end{pmatrix}$, i.e. $\begin{pmatrix} z_2 \\ w_2 \end{pmatrix}  = \begin{pmatrix} 0 \\ 1  \end{pmatrix}$, would be mapped to
\begin{align}
\begin{pmatrix} 0 & -1 \\ 1 & 0 \end{pmatrix} \frac{1}{\epsilon_u}  \begin{pmatrix} e^{i\theta} & -\bar{u}e^{-i\theta} \\ ue^{i\theta} & e^{-i\theta} \end{pmatrix} = \frac{1}{\epsilon_u}  \begin{pmatrix} -ue^{i\theta} & -e^{-i\theta} \\ e^{i\theta} & -\bar{u}e^{-i\theta} \end{pmatrix} \, ,
\end{align}
which corresponds to $\begin{pmatrix} -u \\ 1  \end{pmatrix}  \frac{e^{i\theta}}{\epsilon_u}$, i.e. the second column in \eqref{eq:SU2Act1}.

The action of $e_{1,2,3}$ and $R_1$ on $\vz, \vw$ is thus 
\begin{align}
\begin{pmatrix} \vz \\ \vw \end{pmatrix}  \rightarrow
\begin{pmatrix} e^{i\phi} & 0 \\ 0 & e^{-i\phi} \end{pmatrix}
\begin{pmatrix} e^{i\alpha} \epsilon_u^{-1} & -ue^{i\beta} \epsilon_u^{-1} & ae^{i\gamma} \\ 
u e^{-i\alpha} \epsilon_u^{-1} & e^{-i\beta} \epsilon_u^{-1} & be^{-i\gamma} \end{pmatrix} 
e^{i\theta} \, ,
\end{align}
where $\phi$ is some angle corresponding to the quotienting by $U(1)_Q$.
For infinitesimal $u, a, b$ we get
\begin{align}
\begin{pmatrix} 1 & -u & a \\ 
u & 1 & b \end{pmatrix} & \rightarrow
\begin{pmatrix} e^{i\alpha} & -ue^{i\beta} & ae^{i\gamma} \\ 
u e^{-i\alpha}  & e^{-i\beta}  & be^{-i\gamma} \end{pmatrix} e^{i\theta} \\
&= \begin{pmatrix} e^{i(\alpha+\beta)/2} & 0 \\ 0 & e^{-i(\alpha+\beta)/2} \end{pmatrix} 
\begin{pmatrix} 1 & -ue^{i(\beta-\alpha)} & ae^{i(\gamma-\alpha)} \\ 
u e^{i(\beta-\alpha)}  & 1  & be^{i(\beta-\gamma)} \end{pmatrix} e^{i(\alpha-\beta)/2 + i\theta} \\
&\sim \begin{pmatrix} 1 & -ue^{i(\beta-\alpha)} & ae^{i(\gamma-\alpha)} \\ 
u e^{i(\beta-\alpha)}  & 1  & be^{i(\beta-\gamma)} \end{pmatrix} e^{i(\alpha-\beta)/2 + i\theta} \, .
\end{align}
From this we read off the $U(1)$ weights around this fixed point:
\begin{equation}
\begin{tabular}{l|cccc}
 & $\alpha$ & $\beta$ & $\gamma$ & $\theta$ \\
 \hline 
$u$  & $-1$ & $1$ & $0$ & $0$ \\
$a$  & $-1$ & $0$ & $1$ & $0$ \\
$b$  & $0$ & $1$ & $-1$ & $0$ \\
$\theta$  & $1/2$ & $-1/2$ & $0$ & $1$ \\
\end{tabular}
\end{equation}
For the other point in the same orbit we get the same table but with the weights reversed for $\alpha, \beta, \gamma$ and the same for $\theta$.

Doing a similar analysis around the other two points in \eqref{eq:fps} we get the weights
\begin{equation}
\begin{tabular}{l|cccc}
 & $\alpha$ & $\beta$ & $\gamma$ & $\theta$ \\
 \hline 
$u$  & $1$ & $0$ & $-1$ & $0$ \\
$a$  & $0$ & $1$ & $-1$ & $0$ \\
$b$  & $1$ & $-1$ & $0$ & $0$ \\
$\theta$  & $-1/2$ & $0$ & $1/2$ & $1$ \\
\end{tabular}
\end{equation}
and 
\begin{equation}
\begin{tabular}{l|cccc}
 & $\alpha$ & $\beta$ & $\gamma$ & $\theta$ \\
 \hline 
$u$  & $0$ & $-1$ & $1$ & $0$ \\
$a$  & $1$ & $-1$ & $0$ & $0$ \\
$b$  & $-1$ & $0$ & $1$ & $0$ \\
$\theta$  & $0$ & $1/2$ & $-1/2$ & $1$ \\
\end{tabular}
\end{equation}
together with the versions where the charges of $\alpha, \beta, \gamma$ are reversed.

Let us take a moment to explain how we read of the contribution to the index from these tables.
At a given fixed point the local geometry is $\C^3 \times S^1$ and it turns out that it is enough to consider each of these factors separately.

Assume for a moment that we just had $U(1)$ acting on $\C$ in the usual way. Consider the deformed operator $\db + i \eps \nu \wedge \cdot$, where $\nu$ is the $(0,1)$-component of the 1-form dual to the $U(1)$ vector field and $\eps$ a deformation parameter. The associated complex
\begin{equation}
 0 \rightarrow A^{(0,0)} \rightarrow A^{(0,1)} \rightarrow 0 \, ,
\end{equation}
where $A^{(0,\bullet)}$ denotes $(0,\bullet)$-forms decaying fast enough at infinity, is isomorphic to the original $\db$-complex and one can explicitly compute its cohomology. 
In $H^0$ we find functions of the form $e^{-i\eps |z|^2} z^p$, where $p \geq 0$, {provided} $\Im \eps < 0 $. In $H^1$ we have 1-forms $e^{-i\bar{\eps} |z|^2} \zb^p d \zb$, where $p \geq 1$, provided $\Im \eps > 0$. We see that the index will depend on which direction we deform the operator, i.e. on the sign of $\Im \eps$. Calling the two deformations $\db^\pm$ and rewriting the index in terms of  $U(1)$ representations we get
\begin{equation}
\ind_{U(1)} \left( \left[ \db^\pm  \right] \right)  = \left[ \frac{1}{1-s} \right]^\pm \, ,
\end{equation}
where $s$ is the $U(1)$ coordinate and 
\begin{align}
\left[ \frac{1}{1-s} \right]^+ &= -s^{-1} - s^{-2} - \cdots \, , \\
\left[ \frac{1}{1-s} \right]^- &= 1 + s + s^2 + \cdots  \, .
\end{align}

Back in our main calculation, the local geometry around each fixed point is $\C^3 \times S^1$ and we have figured out how $e_{1,2,3}$ and $R_1$ act on each of these factors. Using the above arguments it is then easy to write down the contributions coming from each of the three $\C$'s. On the $S^1$ factor it can be argued that the symbol restricts to the zero symbol with index
\begin{equation}
\ind \left( [0] \right)  = \delta_1(t) = \sum_{n=-\infty}^\infty t^n \, .
\end{equation}
Let $s_i$ be the coordinates of the $e_i$'s and write $s_{ij}=s_i/s_j$. Let $t$ be the coordinate of $R_1$. We then read off the local contribution to the index from the fixed points from the weight-tables above (remembering that we also have their versions with $\alpha, \beta, \gamma$-charges reversed):
\begin{align}
%
& \left[ \frac{1}{1-s_{21}} \right]^\pm \left[ \frac{1}{1-s_{31}} \right]^\pm\left[ \frac{1}{1-s_{23}} \right]^\pm \delta_1(s_{12} t^2) + \left[ \frac{1}{1-s_{12}} \right]^\pm\left[ \frac{1}{1-s_{13}} \right]^\pm\left[ \frac{1}{1-s_{32}} \right]^\pm \delta_1(s_{21} t^2)
\label{eq:indexContribIa} \\ 
%
& \left[ \frac{1}{1-s_{13}} \right]^\pm\left[ \frac{1}{1-s_{23}} \right]^\pm\left[ \frac{1}{1-s_{12}} \right]^\pm \delta_1(s_{31} t^2) + \left[ \frac{1}{1-s_{31}} \right]^\pm\left[ \frac{1}{1-s_{32}} \right]^\pm\left[ \frac{1}{1-s_{21}} \right]^\pm \delta_1(s_{13} t^2) \label{eq:indexContribIIa}\\ 
%
& \left[ \frac{1}{1-s_{32}} \right]^\pm\left[ \frac{1}{1-s_{12}} \right]^\pm\left[ \frac{1}{1-s_{31}} \right]^\pm \delta_1(s_{23} t^2) + \left[ \frac{1}{1-s_{23}} \right]^\pm\left[ \frac{1}{1-s_{21}} \right]^\pm\left[ \frac{1}{1-s_{13}} \right]^\pm \delta_1(s_{32} t^2)  \label{eq:indexContribIIIa}
\end{align}
Note that the parameter $t$ appears as $t^2$ since the fibre is $SO(3)$ rather than $SU(2)$.

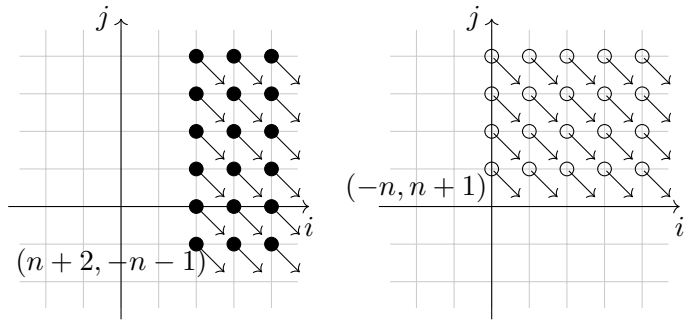
\begin{figure}[!htbp]
\centering
\begin{tikzpicture}[scale=.5]
\draw [step=1,thin,gray!40] (-2.7,-2.7) grid (4.7,4.7);
\draw [->] (-3,0) -- (5,0) node [below] {\small$i$};
\draw [->] (0,-3) -- (0,5) node [left] {\small$j$};
\node at (-0.25,-1.5) {\small$(n+2,-n-1)$};
\foreach \x in {2,...,4}
\foreach \y in {-1,...,4}
	{
	\draw [black,fill=black] (\x,\y) circle (1.0ex);
	\draw [->] (\x,\y) -- (\x+0.75,\y-0.75);
	 }
\end{tikzpicture}
\begin{tikzpicture}[scale=.5]
\draw [step=1,thin,gray!40] (-2.7,-2.7) grid (4.7,4.7);
\draw [->] (-3,0) -- (5,0) node [below] {\small$i$};
\draw [->] (0,-3) -- (0,5) node [left] {\small$j$};
\node at (-2.0,0.5) {\small$(-n,n+1)$};
\foreach \x in {0,...,4}
\foreach \y in {1,...,4}
	{
	\draw [black] (\x,\y) circle (1.0ex);
	\draw [->] (\x+0.05,\y-0.05) -- (\x+0.75,\y-0.75);
	 }
\end{tikzpicture}
\caption{Pictorial representation of the coefficients of $t^{2n}$, $n\geq 0$, coming from the two terms in \eqref{eq:indexContribI}.}
\label{fig:RegI}
\end{figure}

Denote $q_1 = s_1/s_3$ and $q_2 = s_2/s_3$ and let
\begin{equation}
\phi(q)=1+q +q^2+q^3+\cdots \, .
\end{equation}
We assume that $|q_{1,2}|<1$ to get a concrete choice ($\pm$) of regularisation for each $[\frac{1}{1-s_{ij}}]^\pm$-term. Expanding the expressions in \eqref{eq:indexContribIa}-\eqref{eq:indexContribIIIa} into power series gives:
\begin{align}
%
& \left[ -\tfrac{q_1}{q_2} \phi \left(\tfrac{q_1}{q_2}\right)\right]\left[ -q_1 \phi \left(q_1 \right)\right] \left[ \phi \left({q_2}\right)\right]\delta_1 \left(\tfrac{q_1}{q_2} t^2 \right) 
+ \left[ \phi \left( \tfrac{q_1}{q_2} \right) \right]\left[ \phi(q_1) \right]\left[ -q_2 \phi(q_2) \right] \delta_1 \left(\tfrac{q_2}{q_1} t^2 \right) \label{eq:indexContribI} \\ 
& \left[   \phi \left( q_1 \right)\right]\left[  \phi \left(q_2 \right)\right] \left[ \phi  \left( \tfrac{q_1}{q_2} \right) \right]\delta_1 \left(q_1^{-1} t^2 \right) 
+ \left[  -q_1 \phi \left( q_1 \right)\right]\left[ -q_2 \phi \left(q_2 \right)\right] \left[-\tfrac{q_1}{q_2} \phi  \left( \tfrac{q_1}{q_2} \right) \right]\delta_1 \left(q_1 t^2 \right) \label{eq:indexContribII}\\ 
& \left[  -q_2 \phi \left( q_2 \right)\right]\left[  \phi \left( \tfrac{q_1}{q_2}\right)\right] \left[ -q_1 \phi  \left( q_1\right) \right]\delta_1 \left(q_2 t^2 \right) 
+ \left[ \phi \left( q_2 \right)\right]\left[ -\tfrac{q_1}{q_2} \phi \left( \tfrac{q_1}{q_2}\right)\right] \left[ \phi  \left( q_1\right) \right]\delta_1 \left(q_2^{-1} t^2 \right) \label{eq:indexContribIII}
\end{align}
Now let us calculate the coefficients in front of the $t$'s. These will be power series in $q_1$ and $q_2$ with terms of the form $\pm k q_1^i q_2^j$. We will represent such a term as the integer lattice point $(i,j)$ with a solid or hollow dot corresponding to a $+$ or $-$ coefficient respectively.
Consider first the coefficient of $t^{2n}$ for some fixed $n \geq 0$. 
For \eqref{eq:indexContribI} we get the lattices in Figure \ref{fig:RegI}.

Let us take a moment to explain how we obtain the first picture in Figure \ref{fig:RegI}, which corresponds to the first part of \eqref{eq:indexContribI}, i.e.  $\left[ -\tfrac{q_1}{q_2} \phi \left(\tfrac{q_1}{q_2}\right)\right]\left[ -q_1 \phi \left(q_1 \right)\right] \left[ \phi \left({q_2}\right)\right]\delta_1 \left(\tfrac{q_1}{q_2} t^2 \right)$. From the term $\delta_1 \left(\tfrac{q_1}{q_2} t^2 \right)$ we would get the coefficient $+q_1^n q_2^{-n}$. We would draw this as a solid dot at $(n,-n)$.
However, the pre-factors combine to $+q_1^2 q_2^{-1}$ which shifts the dot to $(n+2,-n-1)$.
Next, multiplying by $\phi(q_2) = 1 + q_2 + q_2^2 + \cdots$ and $\phi(q_1) = 1 + q_1 + q_1^2 + \cdots$ means that we should also include the points above and to the right.
Finally, multiplying by $\phi (\frac{q_1}{q_2})=1+q_1 q_2^{-1}+q_1^2 q_2^{-2}+\cdots$ means that we should also include all points south-east of these ones, which we represent by drawing arrows.
This of course intruduces some multiplicities. For example the point $(n+4,-n+1)$ needs to be counted three times, corresponding to the term $+3 q_1^{n+4} q_2^{-n+1}$.

\begin{figure}[!htbp]
\centering
\begin{tikzpicture}[scale=.5]
\draw [step=1,thin,gray!40] (-2.7,-2.7) grid (4.7,4.7);
\draw [->] (-3,0) -- (5,0) node [below] {\small$i$};
\draw [->] (0,-3) -- (0,5) node [left] {\small$j$};
\node at (-0.75,0.5) {\small$(1,n+1)$};
\foreach \x in {1,...,4}
\foreach \y in {1,...,4}
	{
	\draw [black,fill=black] (\x,\y) circle (1.0ex);
	\draw [->] (\x,\y) -- (\x+0.75,\y-0.75);
	 }
\end{tikzpicture}
\begin{tikzpicture}[scale=.5]
\draw [step=1,thin,gray!40] (-2.7,-2.7) grid (4.7,4.7);
\draw [->] (-3,0) -- (5,0) node [below] {\small$i$};
\draw [->] (0,-3) -- (0,5) node [left] {\small$j$};
\node at (-0.75,-1.5) {\small$(1,-n-1)$};
\foreach \x in {1,...,4}
\foreach \y in {-1,...,4}
	{
	\draw [black] (\x,\y) circle (1.0ex);
	\draw [->] (\x+0.05,\y-0.05) -- (\x+0.75,\y-0.75);
	 }
\end{tikzpicture}
\caption{Pictorial representation of the coefficients of $t^{2n}$, $n\geq 0$, coming from the two terms in \eqref{eq:indexContribIII}.}
\label{fig:RegIII}
\end{figure}
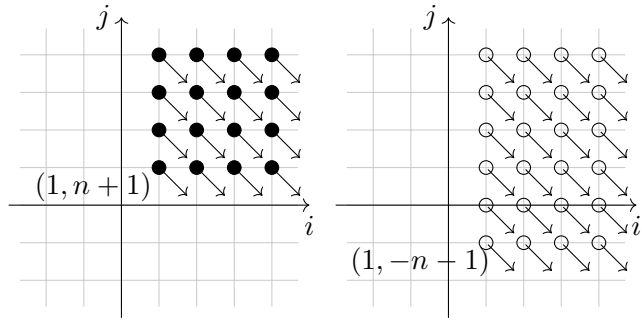

Performing the same analysis for \eqref{eq:indexContribII} and \eqref{eq:indexContribIII} gives Figure \ref{fig:RegII} and \ref{fig:RegIII}.
\begin{figure}[!htbp]
\centering
\begin{tikzpicture}[scale=.5]
\draw [step=1,thin,gray!40] (-2.7,-2.7) grid (4.7,4.7);
\draw [->] (-3,0) -- (5,0) node [below] {\small$i$};
\draw [->] (0,-3) -- (0,5) node [left] {\small$j$};
\node at (-1.25,-0.5) {\small$(-n,0)$};
\foreach \x in {0,...,4}
\foreach \y in {0,...,4}
	{
	\draw [black,fill=black] (\x,\y) circle (1.0ex);
	\draw [->] (\x,\y) -- (\x+0.75,\y-0.75);
	 }
\end{tikzpicture}
\begin{tikzpicture}[scale=.5]
\draw [step=1,thin,gray!40] (-2.7,-2.7) grid (4.7,4.7);
\draw [->] (-3,0) -- (5,0) node [below] {\small$i$};
\draw [->] (0,-3) -- (0,5) node [left] {\small$j$};
\node at (0.5,-0.5) {\small$(n+2,0)$};
\foreach \x in {2,...,4}
\foreach \y in {0,...,4}
	{
	\draw [black] (\x,\y) circle (1.0ex);
	\draw [->] (\x+0.05,\y-0.05) -- (\x+0.75,\y-0.75);
	 }
\end{tikzpicture}
\caption{Pictorial representation of the coefficients of $t^{2n}$, $n\geq 0$, coming from the two terms in \eqref{eq:indexContribII}.}
\label{fig:RegII}
\end{figure}
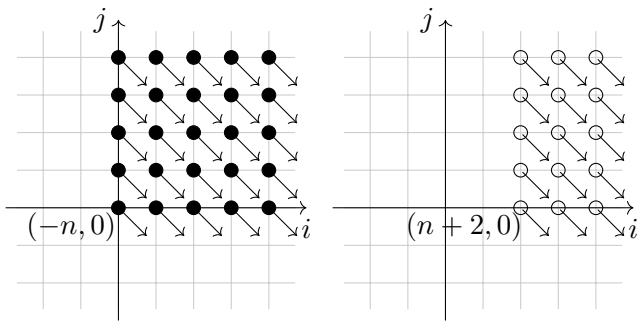

\begin{figure}[!htpb]
\centering
\begin{tikzpicture}[scale=.4]
\draw [step=1,thin,gray!40] (-7.7,-6.7) grid (7.7,5.7);
\draw [->] (-8,0) -- (8,0) node [above] {\small$i$};
\draw [->] (0,-7) -- (0,6) node [left] {\small$j$};
\node at (-5.5,4) {\small$(-n,n)$};
\node at (1.5,4) {\small$(0,n)$};
\node at (-5.5,-0.5) {\small$(-n,0)$};
\foreach \x in {-4,...,0}
\foreach \y in {0,...,4}
	{
	\draw [black,fill=black] (\x,\y) circle (1.0ex);
	\draw [->] (\x,\y) -- (\x+0.75,\y-0.75);
	 }
\node at (-1.5,-5.5) {\small$(1,-n-1)$};
\node at (9.5,-5.5) {\small$(n+1,-n-1)$};
\node at (8,-0.5) {\small$(n+1,-1)$};
\foreach \x in {1,...,5}
\foreach \y in {-5,...,-1}
	{
	\draw [black] (\x,\y) circle (1.0ex);
	\draw [->] (\x,\y) -- (\x+0.75,\y-0.75);
	 }
\end{tikzpicture}
\begin{tikzpicture}[scale=.4]
\draw [step=1,thin,gray!40] (-7.7,-6.7) grid (7.7,5.7);
\draw [->] (-8,0) -- (8,0) node [above] {\small$i$};
\draw [->] (0,-7) -- (0,6) node [left] {\small$j$};
\node at (-5.5,4) {\small$(-n,n)$};
\node at (1.5,4) {\small$(0,n)$};
\node at (-5.5,-0.5) {\small$(-n,0)$};
\node at (-1.5,-4.5) {\small$(0,-n)$};
\node at (5.5,-4.5) {\small$(n,-n)$};
\node at (5.5,0.5) {\small$(n,0)$};
\foreach \n in {0,...,4}
{
	\draw[black] (-\n,\n) -- (-\n,0) -- (0,-\n) -- (\n,-\n) -- (\n,0) -- (0,\n) --cycle;
}
\foreach \x in {-4,...,0}
\foreach \y in {0,...,4}
\foreach \k in {0,...,4}
	{
	\draw [black,fill=black] (\x+\k,\y-\k) circle (1.0ex);
	 }
\end{tikzpicture}
\caption{Net result for the $t^{2n}$-coefficient, $n\geq 0 $, after cancellations.}
\label{fig:RegComb}
\end{figure}
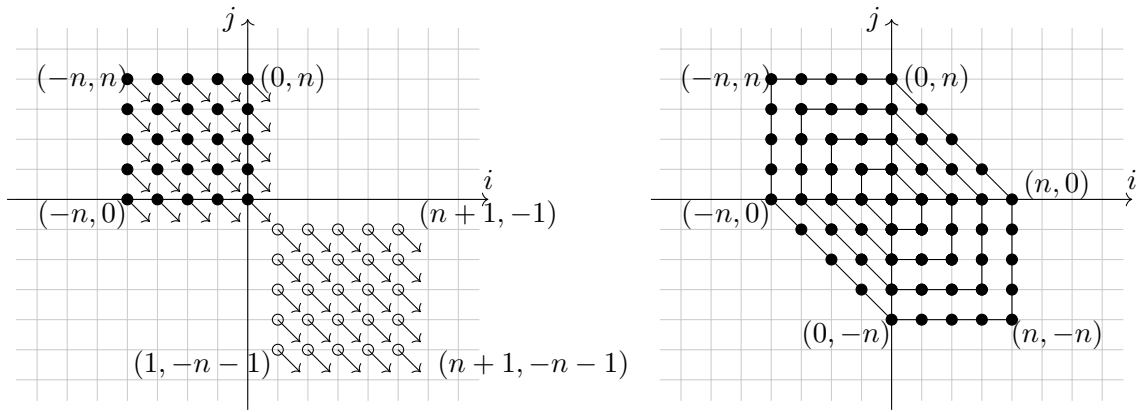

Combining all contributions we get a lot of cancellations. What is left is illustrated in Figure \ref{fig:RegComb}.
The net result is a finite set of points that we can fit inside a polygon. Here the points in the outermost polygon have multiplicity 1, the second outermost multiplicity 2, etc, all the way to the origin which has multiplicity $n+1$. These points with their multiplicities correspond to the following expression for the $t^{2n}$-coefficient:
\begin{equation}
\sum_{i=-n}^0 \sum_{j=0}^n \sum_{k=0}^n q_1^{i+k} q_2^{j-k} \, .
\end{equation}
Thus we can write the contribution to the index from this term as
\begin{equation}
\ind(\db_H)|_{t^{2n}, n\geq 0} = \sum_{i,j,k=0}^n  q_1^{k-i} q_2^{j-k} = \sum_{i,j,k=0}^n  q_1^{-i} q_2^{j} q_3^k \, , \label{eq:indexContribPos}
\end{equation}
where $q_3 = s_1/s_2=q_1/q_2$.

Now we turn to the negative powers, $t^{-2n}$, $n \geq 1$. Firstly, we note that the $t^{-2}$ coefficient is zero. For $n \geq 2$ we just replace $n \leftrightarrow -n$ in the previous analysis. The result is illustrated in Figure \ref{fig:RegCombNeg}.

\begin{figure}[!htbp]
\centering
\begin{tikzpicture}[scale=.5]
\draw [step=1,thin,gray!40] (-5.7,-4.7) grid (5.7,3.7);
\draw [->] (-6,0) -- (6,0) node [above] {\small$i$};
\draw [->] (0,-5) -- (0,4) node [left] {\small$j$};
\node at (-4,-0.5) {\small$(-n+2,0)$};
\node at (-5,2) {\small$(-n+2,n-2)$};
\node at (2.5,2) {\small$(0,n-2)$};
\foreach \x in {-2,...,0}
\foreach \y in {0,...,2}
	{
	\draw [black] (\x,\y) circle (1.0ex);
	\draw [->] (\x,\y) -- (\x+0.75,\y-0.75);
	 }
\node at (-1,-3.5) {\small$(1,-n+1)$};
\node at (7,-3.0) {\small$(n-1,-n+1)$};
\node at (6,-1.0) {\small$(n-1,-1)$};
\foreach \x in {1,...,3}
\foreach \y in {-3,...,-1}
	{
	\draw [black,fill=black] (\x,\y) circle (1.0ex);
	\draw [->] (\x,\y) -- (\x+0.75,\y-0.75);
	 }
\end{tikzpicture}
\begin{tikzpicture}[scale=.5]
\draw [step=1,thin,gray!40] (-5.7,-4.7) grid (5.7,3.7);
\draw [->] (-6,0) -- (6,0) node [above] {\small$i$};
\draw [->] (0,-5) -- (0,4) node [left] {\small$j$};
\node at (-4,-0.5) {\small$(-n+2,0)$};
\node at (-5,2) {\small$(-n+2,n-2)$};
\node at (2.5,2) {\small$(0,n-2)$};
\node at (-1.5,-2.5) {\small$(0,-n)$};
\node at (3.5,-2.5) {\small$(n-2,-n+2)$};
\node at (4,0.5) {\small$(n-2,0)$};
\foreach \n in {0,...,2}
{
	\draw[black] (-\n,\n) -- (-\n,0) -- (0,-\n) -- (\n,-\n) -- (\n,0) -- (0,\n) --cycle;
}
\foreach \x in {-2,...,0}
\foreach \y in {0,...,2}
\foreach \k in {0,...,2}
	{
	\draw [black] (\x+\k,\y-\k) circle (1.0ex);
	 }
\end{tikzpicture}
\caption{Net result for the $t^{-2n}$-coefficient, $n\geq2$, after cancellations.}
\label{fig:RegCombNeg}
\end{figure}

We get the $t^{-2n}$ coefficient
\begin{equation}
-\sum_{i=-n+2}^0 \sum_{j=0}^{n-2} \sum_{k=0}^{n-2} q_1^{i+k} q_2^{j-k} \, , 
\end{equation}
and the contribution to the index is thus
\begin{equation}
\ind(\db_H)|_{t^{-2n}, n\geq 2} = -\sum_{i,j,k=0}^{n-2}  q_1^{k-i} q_2^{j-k} = -\sum_{i,j,k=0}^{n-2}  q_1^{-i} q_2^{j} q_3^k \, . \label{eq:indexContribNeg}
\end{equation}

Let $\w_{1,2,3}, \mu$ be equivariant parameters for $e_{1,2,3}$ and $R_1$, and let $\w_{ab}=\w_a-\w_b$.  We then combine \eqref{eq:indexContribPos} and \eqref{eq:indexContribNeg} to obtain the infinite product expression for the superdeterminant \eqref{eq:sdet}
\begin{equation}
sdet_{\Omega_H^{(0,\bullet)}}(-\L_R + x) \sim 
\frac{\prod\limits_{n=0}^\infty \prod\limits_{i,j,k=0}^n \left( i\w_{31} + j\w_{23} + k\w_{12} + 2n \mu + x\right)} {\prod\limits_{n=2}^\infty \prod\limits_{i,j,k=0}^{n-2} \left( i\w_{31} + j\w_{23} + k\w_{12} - 2 n \mu + x\right)} \, ,
\end{equation}
which we can rewrite as 
\begin{equation}
sdet_{\Omega_H^{(0,\bullet)}}(-\L_R + x) \sim 
\frac{\prod\limits_{n=0}^\infty \prod\limits_{i,j,k=0}^n \left( i\w_{31} + j\w_{23} + k\w_{12} + 2n \mu + x\right)} {\prod\limits_{n=2}^\infty \prod\limits_{i,j,k=0}^{n-2} \left( i\w_{31} + j\w_{23} + k\w_{12} + 2 n \mu - x\right)}
\, . \label{eq:indexAnswer}
\end{equation}
This is in agreement with the expression \eqref{eq:holofunAnswer} obtained via the hypertoric method.

\section{Superdeterminant in terms of $SO(3)$} \label{app:sdetSO3}

Here we will write the result of the superdeterminant calculation in terms of $SO(3)$ data.
The method here is based on the index calculation in appendix \ref{sec:IndexCalc}. 

Let 
\begin{equation}
\chi(q;n) = \sum_{i=-n}^n q^i 
\end{equation}
denote the character of $SU(2)$ at weight $n$, and let 
\begin{equation}
S(q;t) = \sum_{n=0}^\infty \chi(q; n+\tfrac{1}{2})t^{2n} - \sum_{n=2}^\infty \chi(q; n-\tfrac{3}{2})t^{-2n} \label{eq:Sqt}
\end{equation}
be a collection of such characters.

The contributions \eqref{eq:indexContribI}-\eqref{eq:indexContribIII} to the index can be rewritten as
\begin{align}
 -q_2 q_3^{\frac12} \phi(q_2) \phi(q_2q_3) S(q_3;t) \, ,\label{eq:indexContribISU2} \\ 
 q_1^{\frac12} \phi(q_2) \phi(q_1 q_2^{-1}) S(q_1;t) \, , \label{eq:indexContribIISU2}\\ 
  -q_1 q_2^{-\frac12} \phi(q_1) \phi(q_1 q_2^{-1}) S(q_2;t) \, . \label{eq:indexContribIIISU2}
\end{align}
Here the $S(q;t)$ give the contribution from the fibre and the $\phi$'s the contribution from the normal bundle of that fibre.

The $S(q;t)$ in \eqref{eq:Sqt} corresponds to the infinite product expression
\begin{equation}
\frac{\prod\limits_{\substack{k,l=0 \\ k+l=odd}}^{\infty} \left( x- \mu + k(\mu -\frac12 \w) + l(\mu+\frac12 \w) \right) }{\prod\limits_{\substack{k,l=1 \\ k+l=odd}}^{\infty} \left( -(x- \mu) + k(\mu -\frac12 \w) + l(\mu+\frac12 \w) \right) } \, , \label{eq:SqtProd}
\end{equation}
where $\w$ and $\mu$ are the parameters corresponding to $q$ and $t$ respectively.
Recalling the infinite product expression of the double sine function
\begin{equation}
S_2(x|\w_1,\w_2) = \frac{\prod\limits_{\substack{k,l=0}}^{\infty} \left( x + k \w_1 + l\w_2 \right) }{\prod\limits_{\substack{k,l=1}}^{\infty} \left( -x + k \w_1 + l\w_2  \right)} \, ,
\end{equation}
we see that the expression above can be rewritten in terms of $S_2^{odd}$, i.e. $S_2$ restricted to the sublattice of $\mathbb{Z}_{\geq 0}^2$ where $k+l$ is odd. We thus write \eqref{eq:SqtProd} as
\begin{equation}
S_2^{odd} \left(x-\mu|\mu -\tfrac12 \w, \mu +\tfrac12 \w \right) \, .
\end{equation}
Based on this observation we rewrite \eqref{eq:indexContribISU2}-\eqref{eq:indexContribIIISU2} as
\begin{align}
&  \prod_{i=1, j=0} S_2^{odd} \left( x+ i\w_{23}+ j\w_{13} - \mu +\tfrac12 \w_{12} | \mu -\tfrac12 \w_{12}, \mu + \tfrac12 \w_{12}\right)^{-1}  \, ,\label{eq:indexContribIS2Odd} \\ 
&  \prod_{i=0, j=0} S_2^{odd} \left( x+ i\w_{12}+ j\w_{23} - \mu +\tfrac12 \w_{13} | \mu -\tfrac12 \w_{13}, \mu + \tfrac12 \w_{13}\right) \, ,\label{eq:indexContribIIS2Odd}\\ 
&  \prod_{i=1, j=0} S_2^{odd} \left( x+ i\w_{12}+ j\w_{13} - \mu +\tfrac12 \w_{23} | \mu -\tfrac12 \w_{23}, \mu + \tfrac12 \w_{23}\right)^{-1} \, . \label{eq:indexContribIIIS2Odd}
\end{align}
We thus obtain the following expression for the superdeterminant:
\begin{align}
sdet_{\Omega_H^{(0,\bullet)}}(-\L_R + x) \sim 
& \left( \prod_{i=1, j=0} S_2^{odd} \left( x+ i\w_{23}+ j\w_{13} - \mu +\tfrac12 \w_{12} | \mu -\tfrac12 \w_{12}, \mu + \tfrac12 \w_{12}\right)^{-1} \right)  \notag \\ 
&  \left(  \prod_{i=0, j=0} S_2^{odd} \left( x+ i\w_{12}+ j\w_{23} - \mu +\tfrac12 \w_{13} | \mu -\tfrac12 \w_{13}, \mu + \tfrac12 \w_{13}\right) \right) \label{eq:sdetIndexSO3}\\ 
&  \left( \prod_{i=1, j=0} S_2^{odd} \left( x+ i\w_{12}+ j\w_{13} - \mu +\tfrac12 \w_{23} | \mu -\tfrac12 \w_{23}, \mu + \tfrac12 \w_{23}\right)^{-1} \right)  \, . \notag
\end{align}

\section{Calculation of asymptotics} \label{AppAsymp}

Here we perform an analysis of the asymptotic behaviour of the expressions in \eqref{eq:factorisationFunI}. The methods here follow closely those in \cite{Qiu:2013pta}.

Before turning to the functions at hand, let us discuss the asymptotic behaviour of the ordinary quadruple sine to illustrate the methods.

\subsection{Asymptotics of the quadruple sine}

The ordinary quadruple sine function can be represented by the infinite product
\begin{equation}
S_4 (x|\w_1,\w_2,\w_3,\w_4) = \frac{\prod\limits_{n_1,n_2,n_3,n_4 \geq 0}\left( n_1\w_1+n_2\w_2+n_3\w_3+n_4\w_4 + x \right) } {\prod\limits_{n_1,n_2,n_3,n_4 \geq 1}\left( n_1\w_1+n_2\w_2+n_3\w_3+n_4\w_4 - x \right)} \, . \label{eq:S4DefnApp}
\end{equation}
Let us start with the numerator. We zeta-function regularise this product via
\begin{align*}
& -\log \prod \limits_{n_1,n_2,n_3,n_4=0}^\infty (n_1\w_1+n_2\w_2+n_3\w_3+n_4\w_4+x) \notag \\
&= \frac{\partial}{\partial s} \frac{1}{\Gamma(s)}\int_0^\infty \sum_{n_1,n_2,n_3,n_4=0}^\infty e^{-(x+n_1\w_1+n_2\w_2+n_3\w_3+n_4\w_4) t} t^{s-1} dt \Big|_{s=0} \\
& = \frac{\partial}{\partial s} \frac{1}{\Gamma(s)}\int_0^\infty  e^{-xt} \left(\frac{1}{1-e^{-\w_1 t}}\right)\left(\frac{1}{1-e^{-\w_2 t}}\right)\left(\frac{1}{1-e^{-\w_3 t}}\right)\left(\frac{1}{1-e^{-\w_4 t}}\right)   t^{s-1} dt \Big|_{s=0} \, ,
\end{align*}
where we assumed $\Re \w_i > 0$ to do the summations.
As in section 6 of \cite{Qiu:2013pta} we find the large $\Im x$ behaviour by taking the Laurent series of the integrand around $t=0$, truncating it at $\O(t^0)$, and performing the integral.
Then we perform the same analysis for the denominator of  \eqref{eq:S4DefnApp}. This  can be done by just replacing `$x$' with `$\w_1+\w_2+\w_3+\w_4 -x$' in the above analysis. Combining the result from the numerator and denominator one arrives at 
\begin{align}
- \log S_4 (x|\w_1,\w_2,\w_3,\w_4) \sim 
i\pi \sgn(\Im x) \left( b_4 x^4 +b_3 x^3 +b_2 x^2 +b_1 x+ b_0 \right) \, ,
\end{align}
where
\begin{align}
b_4 &= \frac{1}{24 \w_1\w_2\w_3\w_4} \\
b_3 &= -\frac{\w_1+\w_2+\w_3+\w_4}{12 \w_1\w_2\w_3\w_4}\\
b_2 &= \frac{\w_1^2+\w_2^2+\w_3^2+\w_4^2 + 3 \left(\w_1(\w_2+\w_3+\w_4)+\w_2(\w_3+\w_4)+\w_3\w_4 \right)}{24 \w_1\w_2\w_3\w_4} \\
b_1 &=-\frac{\left(\w_1+\w_2+\w_3+\w_4 \right) \left(\w_1(\w_2+\w_3+\w_4)+\w_2(\w_3+\w_4)+\w_3\w_4 \right)}{24 \w_1\w_2\w_3\w_4}  \\
b_0 &= -\frac{1}{720 \w_1 \w_2 \w_3 \w_4} \Big(\w_1^4-5 \w_1^2 \left(\w_2^2+3 \w_2 (\w_3+\w_4)+\w_3^2+3 \w_3 \w_4+\w_4^2\right)  \notag \\
&\quad -15 \w_1 \left(\w_2^2 (\w_3+\w_4)+\w_2 \left(\w_3^2+3 \w_3 \w_4+\w_4^2\right)+\w_3 \w_4 (\w_3+\w_4)\right)+\w_2^4 \\
&\quad -5 \w_2^2 \left(\w_3^2+3 \w_3 \w_4+\w_4^2\right)-15 \w_2 \w_3 \w_4 (\w_3+\w_4)+\w_3^4-5 \w_3^2 \w_4^2+\w_4^4 \Big) \notag
\end{align}
Note that $b_4 x^4 +b_3 x^3 +b_2 x^2 +b_1 x+ b_0=\frac{1}{4!} B_{4,4}(x,\w_1,\w_2\,w_3,\w_4)$, where $B_{4,4}$ is the multiple Bernoulli polynomial defined in \cite{Narukawa}.

An important property of the quadruple sine function is the following factorisation \cite{Narukawa}:
\begin{align} \label{eq:factorS7}
S_4 (x|\w_1,\w_2\,\w_3,\w_4) =  e^{\frac{\pi i }{4!} B_{4,4}(x|\w_1,\w_2\,w_3,\w_4)}\prod_{k =1}^4 (z_k|\qvec_k)_\infty \, ,
\end{align}
where $z_k = e^{2\pi i \frac{x}{\w_k}}$ and $\qvec_k = \left( e^{2\pi i \frac{\w_1}{\w_k}},\dots ,e^{2\pi i \frac{\w_{k-1}}{\w_k}},e^{2\pi i \frac{\w_{k+1}}{\w_k}},\dots,e^{2\pi i \frac{\w_4}{\w_k}}\right)$. The $q$-shifted factorial $(z|\qvec)_\infty$ is defined as follows \cite{Narukawa}:
Let $z=e^{2\pi i \xi}$, $q_j=e^{2\pi i \tau_j}$, where $\xi,\tau_j \in \C$, $\Im(\tau_j) \neq 0$, $j=0,\dots, r$ and denote $\qvec = (q_0,\dots,q_r)$. Assume that $\Im(\tau_j)<0$ for $j=0,\dots,k-1$ and $\Im(\tau_j)>0$ for $j=k,\dots,r$, and define
\begin{equation}
(z|\qvec)_\infty = \prod \limits_{j_0,\cdots,j_r=0}^\infty \left( 1-x q_0^{-j_0-1} \cdots q_{k-1}^{-j_{k-1}-1} q_k^{j_k} \cdots q_r^{j_r}  \right)^{(-1)^k} \,.
\end{equation}
Further imposing symmetry under re-ordering of the $q$'s makes this function defined for all $\tau_j$ with $\Im (\tau_j) \neq 0$.

\subsection{Asymptotics of the expressions in \eqref{eq:factorisationFunI}}

Now let us turn to the asymptotics of the various factors in \eqref{eq:factorisationFunI}. By similar arguments as above we find the asymptotic behaviour
\begin{align}
- \log (\text{LHS of \eqref{eq:factorisationFunI}}) \sim 
i\pi \sgn(\Im x) \left(c_4 x^4 + c_3 x^3 +  c_2 x^2 + c_1 x+ c_0 \right) \, ,
\end{align}
where
\begin{align}
c_4 &= \frac{\left( \mu^2- \frac{1}{24}(\w_{31}^2+\w_{23}^2+\w_{12}^2) \right)}{(4\mu^2-\w_{12}^2)(4\mu^2-\w_{31}^2)(4\mu^2-\w_{23}^2)} \\
c_3 &= \frac{ -8 \mu\left(\mu^2-\frac{1}{24}(\w_{31}^2+\w_{23}^2+\w_{12}^2) \right)}{(4\mu^2-\w_{12}^2)(4\mu^2-\w_{31}^2)(4\mu^2-\w_{23}^2)} 
\\
c_2 &= \frac{24\left( \mu^2-\frac{1}{48}(\w_{31}^2+\w_{23}^2+\w_{12}^2) \right) \left(\mu^2-\frac{1}{24}(\w_{31}^2+\w_{23}^2+\w_{12}^2) \right) }{(4\mu^2-\w_{12}^2)(4\mu^2-\w_{31}^2)(4\mu^2-\w_{23}^2)} \\
c_1 &= \frac{-32 \mu  \left(\mu^2-\frac{1}{24}(\w_{31}^2+\w_{23}^2+\w_{12}^2) \right) \left( \mu^2-\frac{1}{16}(\w_{31}^2+\w_{23}^2+\w_{12}^2) \right)}{(4\mu^2-\w_{12}^2)(4\mu^2-\w_{31}^2)(4\mu^2-\w_{23}^2)} 
\\
c_0 &= \frac{ \frac{ 8\mu^2}{5}\left(\mu^2-\frac{1}{2}(\w_{31}^2+\w_{23}^2+\w_{12}^2) \right) \left(\mu^2-\frac{1}{24}(\w_{31}^2+\w_{23}^2+\w_{12}^2) \right)}{(4\mu^2-\w_{12}^2)(4\mu^2-\w_{31}^2)(4\mu^2-\w_{23}^2)}  \\ 
&\quad +\frac{\frac{24}{5} \left(\mu^2-\frac{1}{24}(\w_{31}^2+\w_{23}^2+\w_{12}^2) \right)^2}{(4\mu^2-\w_{12}^2)(4\mu^2-\w_{31}^2)(4\mu^2-\w_{23}^2)} 
+\frac{1}{120} \notag \, .
\end{align}
It would be interesting to find a geometrical interpretation of these coefficients. In particular the the leading coefficient ought to be related to the volume of the manifold. It would be interesting to investigate this further and also see if the `unsquashed' geometry is singled out via a minimisation of this volume, similar in sprit to the toric Sasaki-Einstein case discussed in \cite{Martelli:2006yb}.

Turning to the RHS of \eqref{eq:factorisationFunI} we perform a similar analysis for each of the $S_2^{odd}$ factors.
For the first factor we get 
\begin{align}
-\log S_2^{ev} & \big( x + i\w_{23} + j\w_{13} - \mu +\tfrac12 \w_{12} | \mu - \tfrac12 \w_{12}, \mu  + \tfrac12 \w_{12} \big)^{-1} \\
& \sim  i\pi \sgn (\Im x) \left(c_2^{I}x^2 +c_1^{I}x +c_0^{I} \right) \notag 
\end{align}
where 
\begin{align}
c_2^{I} &= \frac{-1}{4\mu^2-\w_{12}^2} \\
c_1^{I} &= \frac{4\mu-\w_{12}-2 i \w_{23}+ 2 j \w_{31}}{4\mu^2-\w_{12}^2} \\
c_0^{I} &= \frac{-\frac{14}{3} \mu^2 + 2 \mu (2 i \w_{23}-2 j \w_{31}+\w_{12})- (i \w_{23}-j \w_{31}) (i \w_{23}-j \w_{31}+\w_{12})}{4\mu^2-\w_{12}^2} + \frac{1}{3}
\end{align}

For the second factor we get 
\begin{align}
-\log S_2^{odd} & \left( x+ i\w_{12}+ j\w_{23} - \mu +\tfrac12 \w_{13} | \mu -\tfrac12 \w_{13}, \mu + \tfrac12 \w_{13}\right)  \\
&\sim  i\pi \sgn (\Im x) \left[c_2^{II}x^2 +c_1^{II}x +c_0^{II} \right] \notag
\end{align}
where
\begin{align}
c_2^{II} &= \frac{1}{4\mu^2-\w_{31}^2}\\
c_1^{II} &= \frac{-4\mu-\w_{31}+2 i \w_{12}+ 2 j \w_{23}}{4\mu^2-\w_{31}^2} \\
c_0^{II} &=\frac{\frac{14}{3} \mu^2 -2 \mu (2 i \w_{12}+2 j \w_{23}-\w_{31})+ (i \w_{12}+j \w_{23}) (i \w_{12}+j \w_{23}-\w_{31})}{4\mu^2-\w_{31}^2} -\frac{1}{3}
\end{align}

The third factor gives
\begin{align}
-\log S_2^{odd} & \left( x + i\w_{12} + j\w_{13} - \mu + \tfrac12 \w_{23} | \mu - \tfrac12 \w_{23}, \mu + \tfrac12 \w_{23} \right)^{-1} \\
&\sim i\pi \sgn (\Im x) \left[c_2^{III}x^2 +c_1^{III}x +c_0^{III} \right] \notag
\end{align}
where
\begin{align}
c_2^{III} &= \frac{-1}{4\mu^2-\w_{23}^2} \\
c_1^{III} &= \frac{4 \mu-\w_{23} -2 i \w_{12}+2 j \w_{31}}{4\mu^2-\w_{23}^2} \\
c_0^{III} &= \frac{-\frac{14}{3} \mu^2+2\mu (2 i \w_{12}-2 j \w_{31}+\w_{23})- (i \w_{12}-j \w_{31}) (i \w_{12}-j \w_{31}+\w_{23})}{4\mu^2-\w_{23}^2}+\frac{1}{3}
\end{align}

Unlike the $q$-shifted factorials in \eqref{eq:factorS7} the $S_2^{odd}$-factors in \eqref{eq:factorisationFunI} have non-trivial asymptotics. Therefore we also need to include Bernoulli-type factors for these terms. The proper factorisation of \eqref{eq:factorisationFunI} is thus
\begin{align}
S(x &|\mu,\w_1,\w_2,\w_3) =  \notag \\
& e^{-i\pi \sgn(\Im x) \left(c_4 x^4 + c_3 x^3 +  c_2 x^2 + c_1 x+ c_0 \right)} \times \label{eq:factorisationFunBern}\\
&\quad \left( \prod_{i=1, j=0} e^{i\pi \sgn (\Im x) \left(c_2^{I}x^2 +c_1^{I}x +c_0^{I} \right)} S_2^{odd} \left( x+ i\w_{23}+ j\w_{13} - \mu +\tfrac12 \w_{12} | \mu -\tfrac12 \w_{12}, \mu + \tfrac12 \w_{12}\right)^{-1} \right)  \notag \\ 
&\quad  \left(  \prod_{i=0, j=0} e^{i\pi \sgn (\Im x) \left(c_2^{II}x^2 +c_1^{II}x +c_0^{II} \right)}S_2^{odd} \left( x+ i\w_{12}+ j\w_{23} - \mu +\tfrac12 \w_{13} | \mu -\tfrac12 \w_{13}, \mu + \tfrac12 \w_{13}\right) \right) \notag \\ 
& \quad \left( \prod_{i=1, j=0} e^{i\pi \sgn (\Im x) \left(c_2^{III}x^2 +c_1^{III}x +c_0^{III} \right)} S_2^{odd} \left( x+ i\w_{12}+ j\w_{13} - \mu +\tfrac12 \w_{23} | \mu -\tfrac12 \w_{23}, \mu + \tfrac12 \w_{23}\right)^{-1} \right)  \, , \notag
\end{align}
where the various polynomial coefficients are given above.


\providecommand{\href}[2]{#2}\begingroup\raggedright

\bibliographystyle{utphys}
\bibliography{7dSwann}{}

\endgroup

\end{document}